\documentclass[12pt, draftclsnofoot, onecolumn]{IEEEtran}
\usepackage{cite}
\ifCLASSINFOpdf
\else
  \usepackage[dvips]{graphicx}
  \graphicspath{{../eps/}}
  \DeclareGraphicsExtensions{.eps}
\fi
\usepackage{bm}
\usepackage{array}
\usepackage{multirow}
\usepackage{url}
\usepackage{subfigure}
\usepackage{lipsum}
\usepackage{mathtools}
\usepackage{cuted}
\usepackage{algorithmic}
\usepackage{algorithm}
\usepackage{setspace}

\usepackage{amsmath,amssymb}

\usepackage{etoolbox}
\makeatletter
\patchcmd{\@makecaption}
  {\scshape}
  {}
  {}
  {}
\makeatother

\begin{document}
\title{A Semi-Blind Multiuser SIMO GFDM System in the Presence of CFOs and IQ Imbalances}

\author{Yujie~Liu,~\IEEEmembership{Member,~IEEE,}
Xu~Zhu,~\IEEEmembership{Senior Member,~IEEE,}
Eng~Gee~Lim,~\IEEEmembership{Senior Member,~IEEE,}
Yufei~Jiang,~\IEEEmembership{Member,~IEEE,}
and Yi~Huang,~\IEEEmembership{Senior Member,~IEEE}

\thanks{This work was be presented in part at the IEEE Globecom, 2019 \cite{28}. This work was supported in part by the AI University Research Centre (AI-URC) through XJTLU Key Programme Special Fund (KSF-P-02), and in part by the Natural Science Foundation of Guangdong Province under grant 2018A030313344. (\emph{Corresponding author: Xu Zhu.})}

\thanks{Y. Liu is with the Department of Electrical Engineering and Electronics, University of Liverpool, Liverpool L69 3GJ, U.K., and also with the Department of Electrical and Electronic Engineering, Xi'an Jiaotong-Liverpool University, Suzhou 215123, China (e-mail: yujieliu@liverpool.ac.uk).}
 \thanks{X. Zhu is with the Department of Electrical Engineering and Electronics, University of Liverpool, Liverpool L69 3GJ, U.K., and also with the School of Electronic and Information Engineering, Harbin Institute of Technology, Shenzhen 518055, China (e-mail: xuzhu@liverpool.ac.uk).}
 \thanks{E. G. Lim is with the Department of Electrical and Electronic Engineering, Xi'an Jiaotong-Liverpool University, Suzhou 215123, China (e-mail: enggee.lim@xjtlu.edu.cn).}
\thanks{Y. Jiang is with the School of Electronic and Information Engineering, Harbin Institute of Technology, Shenzhen 518055, China (e-mail: jiangyufei@hit.edu.cn).}
\thanks{Y. Huang is with the Department of Electrical Engineering and Electronics, University of Liverpool, Liverpool L69 3GJ, U.K. (e-mail: yi.huang@liverpool.ac.uk).}}


\maketitle

%
\begin{abstract}
	In this paper, we investigate an open topic of a multiuser single-input-multiple-output (SIMO) generalized frequency division multiplexing (GFDM) system in the presence of carrier frequency offsets (CFOs) and in-phase/quadrature-phase (IQ) imbalances. A low-complexity semi-blind joint estimation scheme of multiple channels, CFOs and IQ imbalances is proposed. By utilizing the subspace approach, CFOs and channels corresponding to $U$ users are first separated into $U$ groups. For each individual user, CFO is extracted by minimizing the smallest eigenvalue whose corresponding eigenvector is utilized to estimate channel blindly. The IQ imbalance parameters are estimated jointly with channel ambiguities by very few pilots. The proposed scheme is feasible for a wider range of receive antennas number and has no constraints on the assignment scheme of subsymbols and subcarriers, modulation type, cyclic prefix length and the number of subsymbols per GFDM symbol. Simulation results show that the proposed scheme significantly outperforms the existing methods in terms of bit error rate, outage probability, mean-square-errors of CFO estimation, channel and IQ imbalance estimation, while at much higher spectral efficiency and lower computational complexity. The Cram\'er-Rao lower bound is derived to verify the effectiveness of the proposed scheme, which is shown to be close to simulation results.
\end{abstract}


\IEEEpeerreviewmaketitle

\section{Introduction}
Generalized frequency division multiplexing (GFDM) \cite{6}, a generalized form of orthogonal frequency division multiplexing (OFDM), has been regarded as a potential waveform for the next generation wireless communications, as it can maintain most of the benefits of OFDM while overcoming its challenges, \emph{e.g.}, large out-of-band (OOB) emission and high peak-to-average-power-ratio (PAPR) \cite{6,aa2}. It is also particularly attractive for low-latency communications, thanks to a shorter cyclic prefix (CP) required. Moreover, owing to its two-dimensional structure in both time and frequency domains, multiple users can share a GFDM symbol by allocating different subcarriers and/or different subsymbols to each individual user, leading to high flexibility. For instance, multiuser GFDM easily turns out to be orthogonal frequency division multiple access (OFDMA), when the number of subsymbols per GFDM symbol is reduced to one.

However, the benefits of GFDM comes at the cost of inter-carrier interference (ICI) and inter-symbol interference (ISI) caused by its nonorthogonality. Thus, it is difficult to estimate multiple channels for multiuser GFDM from the mixture of the received signals of multiple users, unlike OFDMA which enjoys orthogonality between subcarriers. On the other hand, similarly to OFDM, multiuser GFDM is sensitive to radio frequency (RF) impairments, such as carrier frequency offset (CFO) and in-phase/quadrature-phase (IQ) imbalance. CFO is usually caused by the mismatch between local oscillators at transmitter and receiver or a Doppler frequency shift \cite{1}, and worsens the ICI and ISI in GFDM \cite{1,28}. IQ imbalance is often induced by the gain and phase mismatches between the local oscillator signals utilized for down- and up- conversion of I and Q branches, when low-cost direct-conversion receivers are equipped \cite{18,20,19}, and is likely to incur an additional image interference and lead to biased signal estimates. Thus, the aforementioned issues of the estimation of multiple channels, multiple CFOs and multiple IQ imbalances are critical for multiuser GFDM, which however is still an open area in the literature.

\subsection{Related Work}
The research on multiuser GFDM systems is limited in the literature. Lim \emph{et al.} \cite{1} investigated the impact of CFOs on the system performance of multiuser GFDM, and proposed two multiuser interference cancelation schemes by optimizing the weight and filter coefficient to mitigate the impact of CFOs. The joint subcarrier and subsymbol allocation problem was studied to maximize the sum information decoding rate for multiuser GFDM in \cite{aa1}. In \cite{2}, a low-complexity zero-forcing (ZF) receiver was developed to avoid the huge computation caused by the inverse of a large dimensional channel matrix for generalized frequency division multiple access (GFDMA), where all subsymbols share the same subcarrier assignment. A precoding technique was proposed for PAPR reduction in \cite{3} for GFDMA. However, none of the work in \cite{1,2,3,aa1} considered the estimation of CFOs and channels. With a preamble containing two similar Zadoff-Chu training sequences, multiple CFOs and multiple channels were estimated jointly based on the maximum-likelihood criterion for GFDMA in \cite{4}. However, it demands long training sequences to suppress ICI and ISI, and thus resulting in low spectral efficiency; it assumed a only specific frequency spreading GFDM transmitter structure in \cite{5}. Meanwhile, all the aforementioned work in \cite{1,2,3,4} is for GFDMA with subband carrier assignment scheme (CAS) only. In our previous work \cite{28}, multiple CFOs and multiple channels were estimated semi-blindly for GFDMA with generalized CAS. However, all the aforementioned work in \cite{1,2,3,aa1, 28} and \cite{4} does not taken into account IQ imbalances.


For single-user GFDM systems, scattered pilots were exploited for channel estimation in \cite{8}, whose performance however is more susceptible to frequency selective fading. Least-square (LS) based channel estimation was developed for GFDM in \cite{9}, which however requires high training overhead and may suffer an error floor owing to ICI and ISI. The prototype filters in \cite{10,11,12} were designed to allow interference-free pilot assisted channel estimation, which however work only for specific GFDM systems. By properly localizing the pilots in time domain and utilizing the pilots' information from CP, a linear-minimum-mean-squared-error (LMMSE) based parallel ICI and ISI cancelation method was proposed for channel estimation in GFDM \cite{13}, where two subsymbols are utilized as pilots and the CP length should be the same as the subsymbol length, giving rise to reduced spectral efficiency. The authors in \cite{7} showed that GFDM was sensitive to CFO through SIR analysis, and designed the receiver filter to mitigate the impact of CFO. However, neither CFO estimation nor channel estimation was considered in \cite{7}. In \cite{15}, the CP based blind CFO estimation approach was introduced with a limited estimation range, and a preamble assisted CFO estimation approach was proposed by utilizing a pseudo noise (PN) sequence. Time windowing is combined with the synchronization preamble to reduce OOB emission and estimate CFO jointly in \cite{14}. A preamble containing two similar Zaduff-Chu sequences was utilized to achieve low-complexity CFO estimation in \cite{17}, with just a specific frequency spreading GFDM transmitter structure in \cite{5}. Two subsymbols are adopted as pilots for CFO estimation in \cite{14,15,17}, suffering high training overhead as well as ICI and ISI from data symbols. CFO could be blindly determined by the maximum likelihood approach in \cite{16}, whose estimation range however is limited by the number of subsymbols per GFDM symbol. A robust semi-blind CFO and channel estimation scheme was proposed in our previous work \cite{29} for GFDM, whose CFO and channel were estimated by separate training symbols, demanding high training overhead. IQ imbalance was addressed in \cite{18,20,19} for GFDM systems. Nevertheless, its estimation in \cite{18} and \cite{19} is assisted by a whole GFDM symbol, suffering high training overhead. In summary, the aforementioned work on GFDM have two shortcomings namely: a) they were for a single-user system \cite{8,9,10,11,12,13,14,15,16,17,18,19,20,7}, and b) they dealt with only one of the issues of channel estimation \cite{8,9,10,11,12,13}, CFO estimation \cite{14,15,16,17} and IQ imbalance estimation \cite{18,19,20} without considering their impacts on the other. The impact of CFO, IQ imbalance and phase noise was investigated for full-duplex single-user GFDM systems in \cite{aa3} and \cite{aa5}, which however considered none of the estimation of CFO, channel and IQ imbalance.

Most of the existing work on multiuser OFDM and/or OFDMA has considered either CFO estimation or IQ imbalance estimation. A number of CFO estimation approaches were developed for OFDMA in \cite{21,22,23,24,25}. However, the systems in \cite{21} and \cite{22} were based on interleaved CAS only. The blind CFO estimation approach in \cite{23} was applicable for generalized CAS, which however works only under the assumptions of constant modulus constellation and short CP; and requires prohibitively high complexity for exhaustive search. In \cite{24} and \cite{25}, joint estimation of multiple CFOs and multiple channels was studied for OFDMA with generalized CAS. However, the semi-blind approach proposed in \cite{24} demands a multitude of receive antennas as well as high complexity to search for CFOs, and the preamble-assisted scheme developed by Kalman and particle filtering in \cite{25} reduces the spectral efficiency and significantly underperforms its derived Cram\'er-Rao lower bound (CRLB) especially at medium to high signal-to-noise-ratio (SNR). By utilizing the space alternating generalized expectation maximization (SAGE) approach, IQ imbalance was solved in \cite{26} and \cite{27} for OFDMA systems with two-path successive relaying and bit-interleaved coded modulation, respectively. However, the approaches in \cite{26} and \cite{27} are computationally inefficient due to the iterative implementation and the method in \cite{27} is applicable to interleaved OFDMA only. Both CFO and IQ imbalance were considered and estimated for OFDM systems in \cite{30,31,32}, which however consider a single user only. The work in \cite{aa6} took into account the estimation of CFOs, IQ imbalances and channels for multiuser OFDMA and single-carrier frequency division multiple access, which however assumed a very small CFO. It is noteworthy that the aforementioned work for OFDMA \cite{21,22,23,24,25,26,27,aa6} and OFDM \cite{30,31,32} is not applicable to GFDM, owing to the inherent nonorthogonality of GFDM. In contrast, the approaches for GFDM can typically be applied to OFDM based systems, thanks to the high flexibility of GFDM.


In summary, there is limited research on multiuser GFDM in the presence of RF impairments, \emph{e.g.}, only CFO, not IQ imbalance, was considered in \cite{4} and our previous work \cite{28}; the approaches developed for single-user GFDM \cite{8,9,10,11,12,13,14,15,16,17,18,19,20} addressed only one of the issues of channel estimation, CFO estimation and IQ imbalance estimation; the approaches proposed for OFDMA \cite{21,22,23,24,25,26,27,aa6} and OFDM \cite{30,31,32} which enjoy orthogonality among subcarriers are also not applicable to multiuser GFDM due to the inherent ICI and ISI of GFDM. Besides, the existing work on GFDM \cite{1,2,3,4} can work effectively only for specific GFDM systems, \emph{e.g.}, subband GFDMA \cite{1,2,3,4} and specific frequency spreading GFDM \cite{4}. 


\subsection{Contributions}
Motivated by the above open issues, in this paper, we investigate an uplink single-input-multiple-output (SIMO) GFDM system of $U$ users with generalized assignment scheme of subsymbols and subcarriers in the presence of CFOs and IQ imbalances, and propose a low-complexity semi-blind joint multi-CFO, multi-channel and multi-IQ imbalance estimation (JCCIQE) scheme for the system. First, $U$ CFOs and $U$ channels are separated into $U$ groups by user, assisted by a subspace approach. For each individual user, the CFO is extracted by minimizing the smallest eigenvalue whose corresponding eigenvector is utilized to estimate the channel in a blind manner. Finally, the IQ imbalances are estimated jointly with the channel ambiguities by very few pilots. Our contributions are as follows.

\begin{itemize}
	\item To the best of our knowledge, this is the first work to investigate the estimation of channels and multiple RF impairments (CFOs and IQ imbalances) at the same time for a practical GFDM system, where the previous work on multiuser GFDM \cite{1,2,3,4,28}, single-user GFDM \cite{8,9,10,11,12,13,14,15,16,17,18,19,20}, OFDMA \cite{21,22,23,24,25,26,27,aa6} and OFDM \cite{30,31,32} is not applicable, as summarized in the last paragraph of Subsection \uppercase\expandafter{\romannumeral1}-A. In particular, in our previous work \cite{28}, IQ imbalances were not taken into account, and therefore the channel estimation and CFO estimation approaches proposed in \cite{28} are not applicable in the presence of multiple IQ imbalances. The proposed JCCIQE scheme for multiuser GFDM significantly outperforms the existing methods \cite{9,24,15} in terms of bit error rate (BER), outage probability, mean-square-errors (MSEs) of CFO estimation and equivalent channel estimation. The CRLB on MSE of CFO estimation is derived for the first time for multiuser GFDM systems, which is close to simulation results and therefore verify the effectiveness of the proposed JCCIQE scheme. While CRLB analysis was not presented in \cite{28,1,2,3,4}.
	\item The proposed JCCIQE scheme requires a very low training overhead. Channels and CFOs are first estimated blindly, and joint estimation of channel ambiguities and IQ imbalances is conducted in a semi-blind manner, assisted by the same short pilots serving two purposes at the same time. The resulting training overhead is 64-fold lower than that of \cite{9} and \cite{15} which demand a large number of separate pilots for the estimation of multiple CFOs, channels and IQ imbalances. In our previous work \cite{28}, the pilots were used for channel ambiguities estimation only. Also, unlike \cite{37}, the multi-CFO compensation implemented at receiver side avoids the signaling overhead to feed CFOs back to the respective transmitters.
	
	
	\item The proposed JCCIQE scheme is feasible for a wide range of multiuser GFDM systems. It has no constrains on the assignment scheme of subsymbols and subcarriers, modulation type, CP length, and the number of subsymbols, while the existing approaches for GFDM \cite{1,2,3,4,28,15,9} work only under certain system specifications. Extensive simulation results show that JCCIQE is more robust against ICI, ISI and multiuser interference (MUI) than the pilots assisted approaches \cite{9,15} without an error floor. It also allows a wider range of the number of receive antennas than the approach in \cite{24}. It is also more robust against CFO than \cite{15} and \cite{24}.
	
	\item The proposed JCCIQE scheme demands a low complexity. $U$ CFOs, $U$ channels and $2U$ IQ imbalance parameters are separated and estimated individually, decomposing a complex $4U$-dimensional problem into $4U$ low-complexity one-dimensional problems. It achieves a tens-fold complexity reduction over the approach for OFDMA in \cite{24}.
\end{itemize}

\subsection{Organization and Notations}

The rest of this paper is organized as follows. Section \uppercase\expandafter{\romannumeral2} presents the system model. The proposed low-complexity semi-blind JCCIQE scheme is described in Section \uppercase\expandafter{\romannumeral3}. Performance and complexity analysis are given in Section \uppercase\expandafter{\romannumeral4}. Simulation results are demonstrated in Section \uppercase\expandafter{\romannumeral5}. Section \uppercase\expandafter{\romannumeral6} draws the conclusion.

\emph{Notations}: Bold symbols represent vectors/matrices. Superscripts $T$, $*$, $H$ and $\dagger$ respectively denote the transpose, complex conjugate, complex conjugate transpose and pseudo inverse of a vector/matrix. $\textmd{diag}\{\textbf{a}\}$ indicates a diagonal matrix with vector $\textbf{a}$ on its diagonal. $\bm{0}_{M\times N}$ is an $M\times N$ zero matrix. $\parallel \cdot \parallel^2_\textrm{F}$ is the Frobenius norm. $\mathbb{E}\{\}$ denotes the expectation operator. $\otimes$ is the Kronecker product. $\textrm{det}(\textbf{A})$ denotes the determinant of $\textbf{A}$. $\textbf{A}(a:b,c:d)$ indicates the submatrix of $\textbf{A}$ with rows from $a$ to $b$ and columns from $c$ to $d$. $\jmath$ is the basic imaginary unit. $\textrm{circshift}(\cdot)$ is a function to shift array circularly.

\begin{figure*}[!t]
	\centering
	\includegraphics[width=15cm]{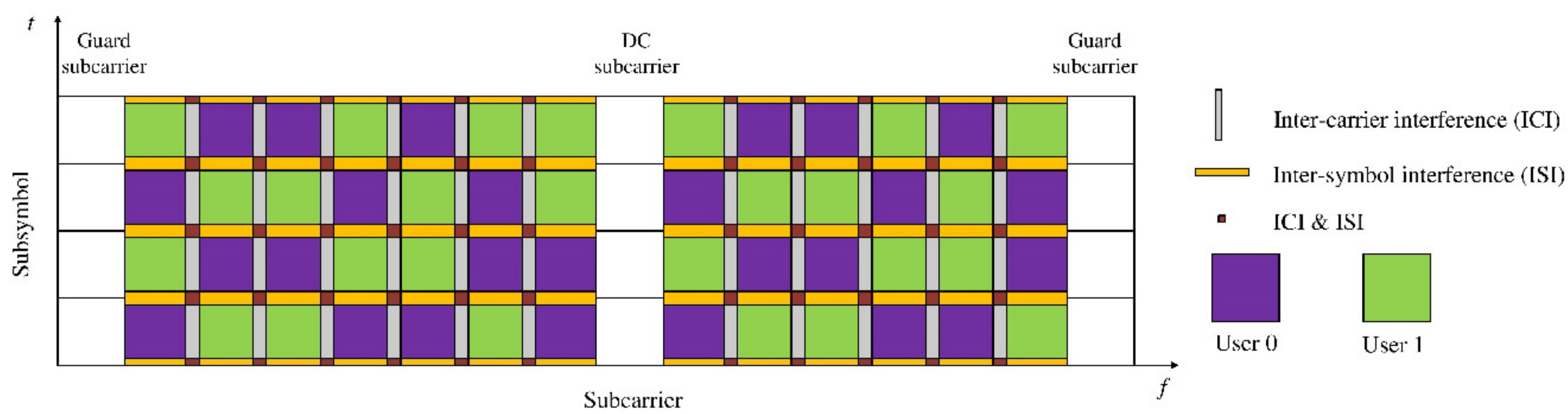}
	\caption{Multiuser GFDM systems with generalized assignment scheme of subsymbols and subcarriers, with $U=2$ users, $M=4$ subsymbols, $K=16$ subcarriers, and $K_{\textrm{D}}=13$ subcarriers.}
	\label{Fig.CAS}
\end{figure*}

\section{System Model}
We consider an uplink $U$-user SIMO GFDM system in the presence of multiple CFOs and multiple IQ imbalances, where each user and the receiver are equipped with a single transmit antenna and $N_{\textrm{r}}$ receive antennas, respectively. Multiple frequency independent (FI) transmit IQ imbalances are considered, assuming the receive IQ imbalance has been compensated \cite{aa6}. 

Each GFDM symbol is divided into $M$ subsymbols each with $K$ subcarriers, and we define $N=MK$. Assume $K_{\textrm{N}}$ subcarriers are utilized as direct current (DC) subcarrier and guard subcarriers, while the rest $K_{\textrm{D}}$ $(K_{\textrm{D}}=K-K_{\textrm{N}})$ subcarriers are used for data transmission. Generalized assignment scheme of subsymbols and subcarriers is considered in this paper. $K_{u,m}$ arbitrary subcarriers on $m$-th subsymbol are assigned to user $u$, with $N_{u}=\sum\nolimits_{m=1}^{M}K_{u,m}$. To guarantee user fairness, $N_{u}$ is assumed to be the same for $u=1,2,\cdots,U$. Define $\mathbb{K}_{u,m}$ as the set of $K_{u,m}$ subcarriers on the $m$-th subsymbol assigned to user $u$, where $\bigcup_{u=1}^{U}\mathbb{K}_{u,m}=\mathbb{K}_{\textrm{D}}$ for $m=1,2,\cdots,M$, and $\mathbb{K}_{u,m}\bigcap\mathbb{K}_{v,m}=\varnothing$, $\forall u\neq v$. $\mathbb{K}_{u,m}(k)$ denotes the $k$-th to the smallest subcarrier index in $\mathbb{K}_{u,m}$, respectively. Fig. \ref{Fig.CAS} illustrates an example of multiuser GFDM systems with generalized assignment scheme of subsymbols and subcarriers, with $U=2$ users, $M=4$ subsymbols, $K=16$ subcarriers, and $K_{\textrm{D}}=13$ subcarriers. Fig. \ref{Fig.CAS} also shows the existence of ICI and ISI in multiuser GFDM systems, which makes its estimation of multiple channels and RF impairments much challenging than that of OFDMA systems. It can be easily noticed that multiuser GFDM systems can easily turn out to be OFDMA systems with generalized CAS in \cite{36}, when the number of subsymbols $M$ decreases to 1.

%


Define $\textbf{d}_{i,u}=[d_{i,u,1,1},\cdots,d_{i,u,1,K_{u,1}},\cdots,d_{i,u,M,1},\cdots,d_{i,u,M,K_{u,M}}]^T$, where $d_{i,u,m,\mathbb{K}_{u,m}(k)}$ is the data of user $u$ in the $m$-th $(m=1,2,\cdots,M)$ subsymbol of $i$-th $(i=1,2,\cdots,N_{\textrm{s}})$ symbol on the $\mathbb{K}_{u,m}(k)$-th $(k=1,2,\cdots,K_{u,m})$ subcarrier, with $N_{\textrm{s}}$ being the number of symbols in a frame. $\textbf{d}_{i,u}$ is transmitted with the corresponding pulse shape \cite{1,2}
\begin{equation}
g_{k,m}[n]=g[(n-mK)\ \textrm{mod} \ N]\cdot \textrm{exp}(-j2\pi kn/K)
\end{equation}
where $n \ (n=1,2,\cdots,N)$ is the sampling index. Note that each $g_{k,m}[n]$ is a time and frequency shifted version of a prototype filter $g_{1,1}[n]$, where the modulo operation performs a circularly shifted version of $g_{k,1}(n)$ and the complex exponential makes $g_{k,m}[n]$ a frequency shifted version of $g_{1,m}[n]$. After pulse shaping, the transmit signal $x_{i,u}[n]$ is given by
\begin{equation}
x_{i,u}[n]=\sum\nolimits_{m=1}^{M}\sum\nolimits_{k=1}^{K_{u,m}}g_{\mathbb{K}_{u,m}(k),m}[n]d_{i,u,m,\mathbb{K}_{u,m}(k)}
\end{equation}
Denote $\textbf{x}_{i,u}=[x_{i,u}[1],x_{i,u}[2],\cdots,x_{i,u}[N]]^T$ as the transmit signal vector of user $u$ in the $i$-th symbol, which can be expressed as $\textbf{x}_{i,u}=\textbf{A}\bm{\Gamma}_u\textbf{d}_{i,u}$
where $\textbf{A}=[\textbf{g}_{1,1},\cdots,\textbf{g}_{K,1},\cdots,\textbf{g}_{1,M},\cdots,$\\
$\textbf{g}_{K,M}]$ is an $N\times N$ pulse shaping filter matrix with $\textbf{g}_{k,m}=[g_{k,m}[1],g_{k,m}[2],\cdots,g_{k,m}[N]]^T$, and $\bm{\Gamma}_u$ is the joint subsymbol and subcarrier assignment matrix of size $N\times N_{u}$ whose $p$-th column vector corresponds to the $p$-th column of the identity matrix $\textbf{I}_{N}$. A CP of length $L_{\textrm{cp}}$ is pre-pended to the symbol $\textbf{x}_{i,u}$, resulting a signal vector $\textbf{s}_{i,u}$ of length $G=N+L_{\textrm{cp}}$, which is given by $\textbf{s}_{i,u}=\bm{\Psi}_u\textbf{d}_{i,u}$
where $\bm{\Psi}_u=\textbf{A}_{\textrm{cp}}\bm{\Gamma_u}$ and $\textbf{A}_{\textrm{cp}}=[\textbf{A}^T(N-L_{\textrm{cp}}+1:N,1:N),\textbf{A}^T]^T$. Note that the modulation matrix $\textbf{A}$ of GFDM is nonorthogonal along subsymbols and subcarriers, unlike the modulation matrix of OFDM which is orthogonal between any two subcarriers. Hence, most previous work in the literature on OFDM based systems cannot be directly extended to GFDM based systems. Note that if there is only one subsymbol in a GFDM symbol, the modulation matrix $\textbf{A}$ would be the Discrete Fourier Transform (DFT) matrix of OFDM.


Define $\theta_u$ and $\epsilon_u$ as the phase and amplitude mismatches between the I and Q branches of user $u$. The asymmetric IQ imbalance model in \cite{bb1} and \cite{bb2} is considered in this paper. Note that the symmetric model can be easily obtained from the asymmetric model through some manipulations, \emph{e.g.}, by means of a rotation matrix and a scaling factor \cite{bb3}. Thus, the IQ imbalance parameters $\alpha_u$ and $\beta_u$ of user $u$ are defined as \cite{bb1,bb2}
\begin{equation}
\alpha_u=(1+\epsilon_u e^{j\theta_u})/2, \quad \beta_u=(1-\epsilon_u e^{j\theta_u})/2
\label{eq23}
\end{equation}
The transmit signal $\textbf{s}_{\textrm{IQ},i,u}$ is written as $\textbf{s}_{\textrm{IQ},i,u}=\alpha_{u}\textbf{s}_{i,u}+\beta_u\textbf{s}_{i,u}^*$.


The channel impulse response (CIR) is assumed to remain constant over a frame. Denote
$\pmb{\hbar}_u^{n_{\textrm{r}}}=[\hbar_u^{n_{\textrm{r}}}[1],\cdots,\hbar_u^{n_{\textrm{r}}}[L]]^T$ as the CIR vector for the $n_{\textrm{r}}$-th receive antenna of user $u$, with $L$ being the length of CIR. $\phi_u$ is defined as the normalized CFO between the user $u$ and the receiver with respect to the subcarrier spacing. The CFO range of $[-0.5,0.5)$ in \cite{24} is considered in this paper.
By incorporating the CFO into the channel of each user as in \cite{44}, the time-domain received signal in the $i$-th symbol at the $n_{\textrm{r}}$-th receive antenna can be written as
\begin{equation}
y_i^{n_{\textrm{r}}}[g]=\sum\nolimits_{u=1}^{U}\sum\nolimits_{l=1}^{L}h_u^{n_{\textrm{r}}}[l]e^{j2\pi\phi_u (g-l)/K}s_{\textrm{IQ},i,u}[g-l]+w_i^{n_{\textrm{r}}}[g]
\end{equation}
where $h_u^{n_{\textrm{r}}}[l]=e^{j2\pi\phi_u l/K}\hbar_u^{n_{\textrm{r}}}[l]$ is the CFO-included channel and $w_i^{n_{\textrm{r}}}[g]$ $(g=1,2,\cdots,G)$ is the additive white Gaussian noise with zero mean and variance $\sigma^2$.

The first $(L-1)$ signal samples, $y_i^{n_{\textrm{r}}}[1]$ to $y_i^{n_{\textrm{r}}}[L-1]$, which suffer from ISI, are discarded and are not utilized for estimation \cite{44}. Collecting the received signal samples $y_i^{n_{\textrm{r}}}[L]$ to $y_i^{n_{\textrm{r}}}[G]$ from $N_{\textrm{r}}$ received antennas into a vector, we obtain $\textbf{y}_i=[y_i^{1}[L],\cdots,y_i^{N_{\textrm{r}}}[L],\cdots,y_i^{1}[G],\cdots,y_i^{N_{\textrm{r}}}[G]]^T$ of size $N_{\textrm{r}}(G-L+1)$, which can be expressed as
\begin{equation}
\textbf{y}_i=\sum\nolimits_{u=1}^{U}(\alpha_u\underbrace{\textbf{H}_u\textbf{E}(\phi_u)\bm{\Psi}_u}_{\textbf{G}_{\textrm{I},u}}\textbf{d}_{i,u}+\beta_u\underbrace{\textbf{H}_u\textbf{E}(\phi_u)\bm{\Psi}_{u}^*}_{\textbf{G}_{\textrm{Q},u}}\textbf{d}_{i,u}^*)+\textbf{w}_i
\label{eq01}
\end{equation}
where $\textbf{E}(\phi_u)=\textrm{diag}\{[1,e^{j2\pi\phi_u/K},\cdots,e^{j2\pi\phi_u(G-1)/K}]\}$ is the CFO matrix of user $u$, $\textbf{H}_u$ is the channel circulant matrix expressed as
\begin{equation}
\textbf{H}_u=\begin{bmatrix}\textbf{h}_u(L)&\cdots&\textbf{h}_u(1)&\cdots&\cdots&\bm{0}_{N_{\textrm{r}\times 1}}\\ \vdots&\ddots&\ddots&\ddots&\ddots&\vdots\\\bm{0}_{N_{\textrm{r}\times 1}}&\cdots&\cdots&\textbf{h}_u(L)&\cdots&\textbf{h}_u(1)\end{bmatrix}
\label{eq00001}
\end{equation}
with $\textbf{h}_u(l)=[h^1_u[l],\cdots,h_u^{N_{\textrm{r}}}[l]]^T$, and similarly $\textbf{w}_i=[w_i^{1}[L],\cdots,w_i^{N_{\textrm{r}}}[L],\cdots,w_i^{1}[G],\cdots,w_i^{N_{\textrm{r}}}[G]]^T$.

\section{Low-Complexity Semi-Blind JCCIQE Scheme}

\begin{figure*}[!t]
	\centering
	\includegraphics[width=15cm]{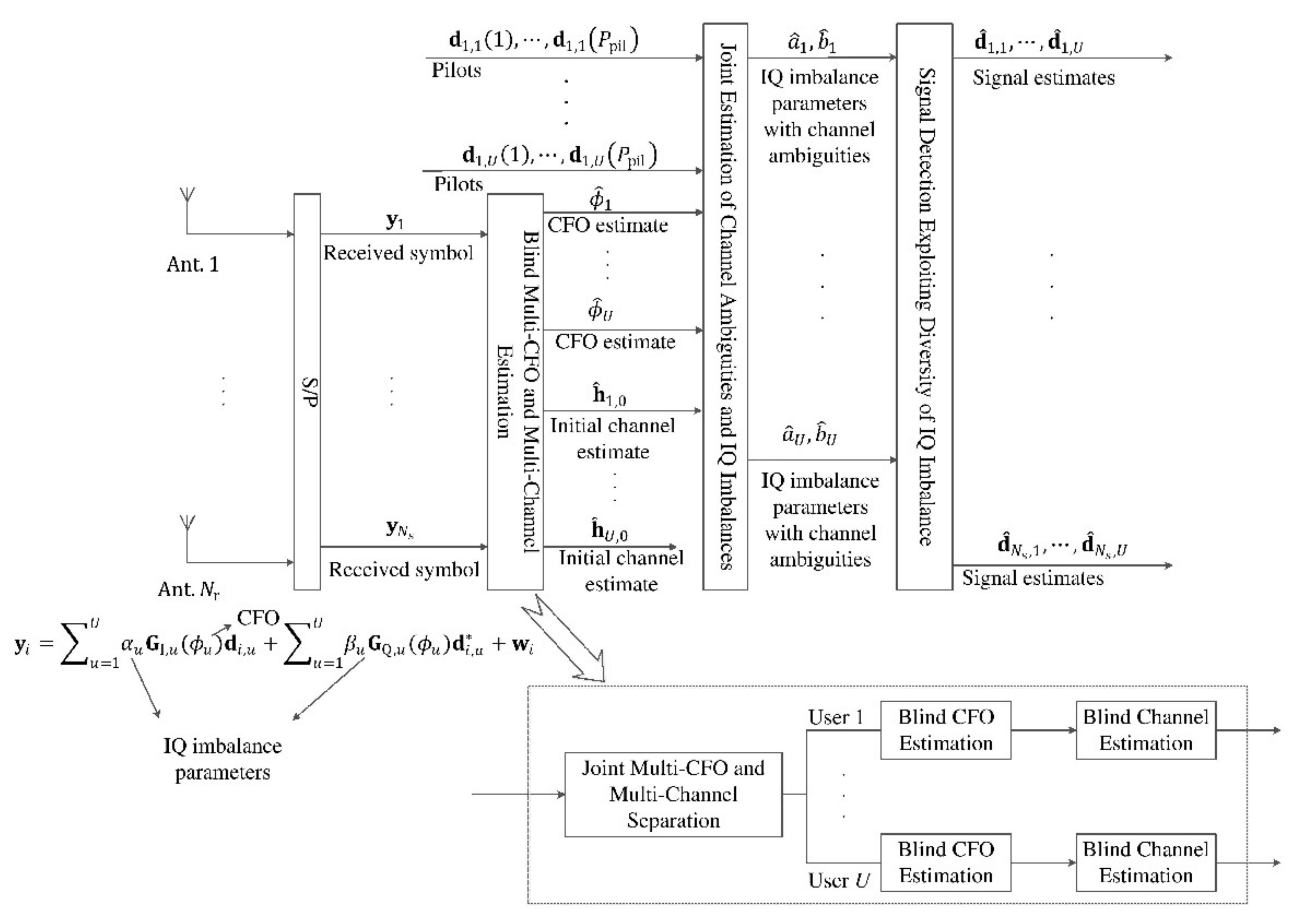}
	\caption{Block diagram of the proposed JCCIQE scheme for multiuser GFDM systems in the presence of multiple CFOs and IQ imbalances (S/P: serial-to-parallel).}
	\label{Fig.BD}
\end{figure*}

The purpose of this work is to determine $U$ CFOs, $U$ channels and $2U$ IQ imbalance parameters from \eqref{eq01}, which is a complex estimation problem of $4U$ dimensions. We propose a low-complexity semi-blind JCCIQE scheme for SIMO multiuser GFDM systems with generalized assignment scheme of subsymbols and subcarriers. Thanks to the orthogonality between noise subspace and each signal subspace of $U$ users, CFOs and channels are first separated into $U$ groups corresponding to $U$ users. Second, for each user, CFO is first estimated blindly by minimizing the smallest eigenvalue to satisfy the orthogonality, while channel is determined blindly by the corresponding eigenvector. Third, for each user, its IQ imbalance parameters and channel ambiguities are estimated jointly by very few pilots. Hence, a $4U$-dimensional estimation problem is decomposed into $4U$ low-complexity one-dimensional estimation problems. Fig. \ref{Fig.BD} illustrates the block diagram of the proposed JCCIQE scheme for multiuser GFDM systems. 

\subsection{Joint Multi-CFO and Multi-Channel Separation}
To separate $U$ channels and $U$ CFOs into $U$ groups by user, the eigenvectors corresponding to noise subspace should be obtained. By exploiting $N_{\textrm{r}}$ receive antennas, the frame with $N_{\textrm{s}}$ received symbols is used to compute the auto-correlation matrix of the received signal, $\textbf{R}_{\textrm{y}}$, which is given by $\textbf{R}_{\textrm{y}}=\frac{1}{N_{\textrm{s}}}\sum\nolimits_{i=1}^{N_{\textrm{s}}}\textbf{y}_i\textbf{y}_i^H$. 

Then, the numbers of noise eigenvectors and signal eigenvectors should be obtained.
For each subsymbol of user $u$, we check if $\mathbb{K}_{u,m}$ intersects with $\mathbb{K}_{\textrm{I},u,m}$, where $\mathbb{K}_{\textrm{I},u,m}=[K-\mathbb{K}_{u,m}+2]_{K}$. Define $\mathbb{I}_{u,m}=\mathbb{K}_{u,m}\bigcap \mathbb{K}_{\textrm{I},u,m}$ and $K_{\textrm{I},u,m}$ as the intersection set and the number of elements in the intersection set, respectively. The number of signal eigenvectors can be calculated by $N_{\textrm{Signal}}=2MK_{\textrm{D}}-\sum\nolimits_{u=1}^{U}\sum\nolimits_{m=1}^{M}K_{\textrm{I},u,m}$.
The value of $N_{\textrm{Signal}}$ varies with the assignment of subsymbols and subcarriers, and is typically smaller than $2MK_{\textrm{D}}$ and larger than $MK_{\textrm{D}}$, \emph{i.e.}, $MK_{\textrm{D}}<N_{\textrm{Signal}}<2MK_{\textrm{D}}$, due to the transmit IQ imbalances. The number of noise eigenvectors is obtained as $Q=N_{\textrm{r}}(G-L+1)-N_{\textrm{Signal}}$, corresponding to the $Q$ smallest eigenvalues of $\textbf{R}_{\textrm{y}}$. Thus, the number of receive antennas should satisfy $N_{\textrm{r}}>\frac{N_{\textrm{Signal}}}{G-L+1}$. Note that given a special case in the absence of IQ imbalances, the numbers of signal and noise eigenvectors are constant, with $N_{\textrm{Signal}}=MK_{\textrm{D}}$ and $Q=N_{\textrm{r}}(G-L+1)-MK_{\textrm{D}}$. 

The $q$-th $(q=1,2,\cdots,Q)$ noise eigenvector is expressed as $\bm{\gamma}_q=[\bm{\gamma}_q^T(1),\bm{\gamma}_q^T(2),\cdots,\bm{\gamma}_q^T(G-L+1)]^T$, where $\bm{\gamma}_q(g)$ is a column vector of length $N_{\textrm{r}}$. Due to orthogonality between the noise subspace and each signal subspace, the columns of $\textbf{G}_{\textrm{I},u}$ and $\textbf{G}_{\textrm{Q},u}$ are orthogonal to each noise eigenvector, \emph{i.e.}, $\bm{\gamma}_q^H\textbf{G}_{\textrm{I},u}=\bm{0}_{1\times N_{u}}$ and $\bm{\gamma}_q^H\textbf{G}_{\textrm{Q},u}=\bm{0}_{1\times N_u}$.
$\textbf{G}_{\textrm{I},u}$ and $\textbf{G}_{\textrm{Q},u}$ differ between different users, thanks to the exclusive assignment of subsymbol and subcarrier for each user. Define $\textbf{h}_u=[\textbf{h}_u^T(L),\textbf{h}_u^T(L-1),\cdots,\textbf{h}_u^T(1)]^T$ as the CFO-included CIR of user $u$. The CFO and the CFO-included CIR of user $u$ can be estimated by
\begin{equation}
[\hat{\phi}_u,\hat{\textbf{h}}_u]=\textrm{arg }\min_{\phi_u,\textbf{h}_u}\sum\nolimits_{q=1}^{Q} (\parallel \bm{\gamma}_q^H\textbf{G}_{\textrm{I},u}\textbf{G}_{\textrm{I},u}^H\bm{\gamma}_q\parallel_{\textrm{F}}^2+\parallel \bm{\gamma}_q^H\textbf{G}_{\textrm{Q},u}\textbf{G}_{\textrm{Q},u}^H\bm{\gamma}_q\parallel_{\textrm{F}}^2)
\label{eq0001}
\end{equation}
Note that $\textbf{G}_{\textrm{Q},u}$ and $\textbf{G}_{\textrm{Q},u}^{H}$ originate from the image signal of user $u$ which usually has lower power and thus lower SNR than the source signal \cite{32}. Thus, the term $\parallel \bm{\gamma}_q^H\textbf{G}_{\textrm{Q},u}\textbf{G}_{\textrm{Q},u}^H\bm{\gamma}_q\parallel_{\textrm{F}}^2$ in \eqref{eq0001} is likely to deteriorate the estimation performance, and can be removed from \eqref{eq0001}, resulting in the following problem formulation:
\begin{equation}
[\hat{\phi}_u,\hat{\textbf{h}}_u]=\textrm{arg }\min_{\phi_u,\textbf{h}_u}\sum\nolimits_{q=1}^{Q} \parallel \bm{\gamma}_q^H\textbf{G}_{\textrm{I},u}\textbf{G}_{\textrm{I},u}^H\bm{\gamma}_q\parallel_{\textrm{F}}^2
\end{equation}

Therefore, by utilizing the orthogonality property, the CFOs and channels can be separated by user. It is noteworthy that the proposed JCCIQE scheme has a looser requirement on the CP length and the number of receive antennas. The CP length takes any value, while it is limited to $L_{\textrm{cp}}<\frac{P-1}{2}$ in \cite{23}, where $P$ is the number of subcarriers per user. The proposed JCCIQE scheme requires a minimum number of receive antennas to satisfy $N_{\textrm{r}}>\frac{N_{\textrm{Signal}}}{KM+L_{\textrm{cp}}-L+1}$ which is smaller than that required in \cite{24}, where the number of receive antennas should satisfy $N_{\textrm{r}}\geq UL$. With $K=16$, $M=4$, $L=3$, $L_{\textrm{cp}}=4$ and $U=4$, the minimum number of receive antennas required by the proposed JCCIQE scheme and the approach in \cite{24} is $N_{\textrm{r}}=2$ and $N_{\textrm{r}}=12$, respectively.

\subsection{Blind Estimation of CFO and Channel for Each User}
After the separation of CFOs and channels by user, each CFO is first extracted to satisfy the orthogonality between noise subspace and each signal subspace by minimizing the smallest eigenvalue, whose corresponding eigenvector is exploited to determine the CFO-included CIR in a blind manner.

Define $\textbf{R}_{u}(\tilde{\phi}_u)=\parallel \bm{\gamma}_q^H\textbf{G}_{\textrm{I},u}(\tilde{\phi}_u)\textbf{G}_{\textrm{I},u}^H(\tilde{\phi}_u)\bm{\gamma}_q\parallel_{\textrm{F}}^2$, where $\textbf{G}_{\textrm{I},u}(\tilde{\phi}_u)=\textbf{H}_u\textbf{E}(\tilde{\phi}_u)\bm{\Psi}_u$, with $\tilde{\phi}_u$ denoting as the CFO trial value of user $u$. $\textbf{R}_{u}(\tilde{\phi}_u)$ can be written as
\begin{equation}
\textbf{R}_u(\tilde{\phi}_u)=\parallel\bm{\gamma}_q^H\textbf{H}_u\textbf{E}(\tilde{\phi}_u)\bm{\Psi}_u\bm{\Psi}_u^H\textbf{E}^H(\tilde{\phi}_u)\textbf{H}_u^H\bm{\gamma}_q\parallel^2_{\textrm{F}}
\label{eq0002}
\end{equation}
According to \cite{40}, $\textbf{H}_u$, as a Toeplitz matrix, has a property of $\bm{\gamma}_q^H\textbf{H}_u=\textbf{h}_u^T\bm{\Upsilon}_q$,
where $\bm{\Upsilon}_q$ is an $N_{\textrm{r}}L\times G$ matrix given by
\begin{equation}
\bm{\Upsilon}_q=\begin{bmatrix}\bm{\gamma}^*_q(1)&\cdots&\bm{\gamma}^*_q(G-L+1)&\cdots&\cdots&\bm{0}_{N_{\textrm{r}\times 1}}\\ \vdots&\ddots&\ddots&\ddots&\ddots&\vdots\\\bm{0}_{N_{\textrm{r}\times 1}}&\cdots&\cdots&\bm{\gamma}^*_q(1)&\cdots&\bm{\gamma}^*_q(G-L+1)\end{bmatrix}
\end{equation}
Therefore, \eqref{eq0002} can be rewritten as
\begin{equation}
\textbf{R}_{u}(\tilde{\phi}_u)=\parallel\textbf{h}_u^T\underbrace{\bm{\Upsilon}_q\textbf{E}(\tilde{\phi}_u)\bm{\Psi}_u}_{\textbf{P}_{u,q}(\tilde{\phi}_u)}\underbrace{\bm{\Psi}_u^H\textbf{E}^H(\tilde{\phi}_u)\bm{\Upsilon}_q^H}_{\textbf{P}_{u,q}^H(\tilde{\phi}_u)}\textbf{h}_u^*\parallel^2_{\textrm{F}}
\end{equation}
Define $\textbf{P}_u(\tilde{\phi}_u)=[\textbf{P}_{u,1}(\tilde{\phi}_u),\textbf{P}_{u,2}(\tilde{\phi}_u),\cdots,\textbf{P}_{u,Q}(\tilde{\phi}_u)]$ of size $N_{\textrm{r}}L\times QN_{u}$. The CFO and CFO-included CIR of user $u$ can be estimated by
\begin{equation}
[\hat{\phi}_u,\hat{\textbf{h}}_u]=\textrm{arg }\min_{\tilde{\phi}_u,\textbf{h}_u} \parallel \textbf{h}_u^T\textbf{P}_u(\tilde{\phi}_u)\textbf{P}_u^H(\tilde{\phi}_u)\textbf{h}_u^*\parallel_{\textrm{F}}^2
\label{eq0003}
\end{equation}
In the following, they are estimated in two steps:

\textbf{Step 1:} The auto-correlation matrix of $\textbf{P}_u(\tilde{\phi}_u)$ is calculated by $\textbf{R}_{\textrm{\textbf{P}},u}(\tilde{\phi}_u)=\textbf{P}_u(\tilde{\phi}_u)\textbf{P}_u^H(\tilde{\phi}_u)$.
If $\tilde{\phi}_u$ is the true CFO $(\tilde{\phi}_u=\phi_u)$, the rank of $\textbf{R}_{\textrm{\textbf{P}},u}(\tilde{\phi}_u)$ is $(N_{\textrm{r}}L-1)$, otherwise the rank is $N_{\textrm{r}}L$. Hence, the CFO of user $u$ can be extracted by minimizing the smallest eigenvalue of $\textbf{R}_{\textrm{\textbf{P}},u}(\tilde{\phi}_u)$, \emph{i.e.}, by minimizing the determinant of $\textbf{R}_{\textrm{\textbf{P}},u}(\tilde{\phi}_u)$
\begin{equation}
\hat{\phi}_u=\textrm{arg }\min_{\tilde{\phi}_u\in [-0.5,0.5)}\textrm{det }(\textbf{R}_{\textrm{\textbf{P}},u}(\tilde{\phi}_u))
\label{eq04}
\end{equation}

Note that to estimate CFO by \eqref{eq04} requires a relatively small search step size for a high accuracy performance, which however demands very high computational complexity. Inspired by \cite{41}, we solve the CFO estimation problem in \eqref{eq04} with a reduced computational complexity by the following two steps:

\textbf{Step 1.1} - Coarse Estimation: The CFO of user $u$ is extracted as $\hat{\phi}_{u,0}$ by \eqref{eq04} with a relatively large search step size $\delta$ to obtain an initial value to approach the global minimum of $\textrm{det}(\textbf{R}_{\textrm{\textbf{P}},u}(\tilde{\phi}_u))$;

\textbf{Step 1.2} - Fine Estimation: A refined CFO is estimated by searching in the range of $(\hat{\phi}_{u,0}-\delta/2,\hat{\phi}_{u,0}+\delta/2)$ by the golden section search and parabolic interpolation algorithms.

\textbf{Step 2:} With the CFO estimate $\hat{\phi}_u$, the CFO-included CIR of user $u$ is determined as the conjugate of the eigenvector corresponding to its smallest eigenvalue of $\textbf{R}_{\textrm{\textbf{P}},u}(\hat{\phi}_u)$, and is denoted as $\hat{\textbf{h}}_{u,0}$.

Note that CFO and CFO-included channel of each user are estimated blindly, which do not suffer the ICI and ISI from data symbols, unlike the existing pilot assisted approaches \cite{9,14,15}. The orthogonality between the signal and noise subspaces is independent of the value of $M$, which allows the CFO estimation in the range of $[-0.5,0.5)$, unlike $[-0.5/M,0.5/M)$ in the CP \cite{15} and maximum likelihood \cite{16} based approaches. As $U$ channels and $U$ CFOs are estimated individually, signaling overhead due to feedback of multiple CFO estimates to transmitters for multi-CFO compensation can be avoided, unlike \cite{37}. It is noteworthy that a complex channel scaling ambiguity $c_u$ exists between the CFO-included CIR estimate $\hat{\textbf{h}}_{u,0}$ and its true CFO-included CIR $\textbf{h}_{u}$, which can be estimated jointly with the IQ imbalance parameters $\alpha_u$ and $\beta_u$ by very few pilots as described in Subsection \uppercase\expandafter{\romannumeral3}-C.

\subsection{Joint Estimation of Channel Ambiguities and IQ Imbalance Parameters for Each User}
With the blind estimates of CFO and CFO-included CIR namely $\hat{\phi}$ and $\hat{\textbf{h}}_{u,0}$ in Subsection \uppercase\expandafter{\romannumeral3}-B, the estimate of matrix $\textbf{G}_{\textrm{I},u}$ and $\textbf{G}_{\textrm{Q},u}$ in \eqref{eq00001} can be denoted as $\hat{\textbf{G}}_{\textrm{I},u}$ and $\hat{\textbf{G}}_{\textrm{Q},u}$, respectively. Assuming a number of subcarriers in the first received GFDM symbol are occupied for pilots, the received signal $\textbf{y}_1$ in \eqref{eq00001} can be rewritten as
\begin{equation}
\textbf{y}_1=\sum\nolimits_{u=1}^{U}\hat{\textbf{G}}_{\textrm{I},u}\textbf{d}_{1,u}\underbrace{c_u\alpha_u}_{a_u}+\sum\nolimits_{u=1}^{U}\hat{\textbf{G}}_{\textrm{Q},u}\textbf{d}_{1,u}^{*}\underbrace{c_u^{*}\beta_u}_{b_u}+\textbf{w}_1
\label{eqaaa}
\end{equation}
where $a_u=c_u\alpha_u$ and $b_u=c_u^*\beta_u$ correspond to the IQ imbalance parameters by taking into account channel ambiguities $c_u$ and $c_u^*$.

Define $\hat{\textbf{G}}=[\hat{\textbf{G}}_{\textrm{I},1},\hat{\textbf{G}}_{\textrm{Q},1},\cdots,\hat{\textbf{G}}_{\textrm{I},U},\hat{\textbf{G}}_{\textrm{Q},U}]$ of size $N_{\textrm{r}}(G-L+1)\times 2MK_{\textrm{D}}$. Due to the intersection set $\mathbb{I}_{u,m}$ in Subsection \uppercase\expandafter{\romannumeral3}-A, $\hat{\textbf{G}}\hat{\textbf{G}}^H$ may be rank deficient, and thus the channel ambiguities along with the IQ imbalance parameters cannot be estimated by ZF algorithm. To address this problem, we carefully place a number of nulls on the subcarriers in the intersection set $\mathbb{I}_{u,m}$ of the first GFDM symbol, \emph{i.e.}, $d_{1,u,m,k}=0$ for $k\in\mathbb{I}_{u,m}$. As a result, we can have
\begin{equation}
\textbf{y}_1=\sum\nolimits_{u=1}^{U}\bar{\textbf{G}}_{\textrm{I},u}\bar{\textbf{d}}_{1,u}a_u+\sum\nolimits_{u=1}^{U}\bar{\textbf{G}}_{\textrm{Q},u}\bar{\textbf{d}}_{1,u}^{*}b_u+\textbf{w}_1
\label{eqbbb}
\end{equation}
where $\bar{\textbf{G}}_{\textrm{I},u}$, $\bar{\textbf{G}}_{\textrm{Q},u}$ and $\bar{\textbf{d}}_{1,u}$ are the version of $\hat{\textbf{G}}_{\textrm{I},u}$, $\hat{\textbf{G}}_{\textrm{Q},u}$ and $\textbf{d}_{1,u}$ with reduced size, by excluding the elements corresponding to nulls.

Similarly, define $\bar{\textbf{G}}=[\bar{\textbf{G}}_{\textrm{I},1},\bar{\textbf{G}}_{\textrm{Q},1},\cdots,\bar{\textbf{G}}_{\textrm{I},U},\bar{\textbf{G}}_{\textrm{Q},U}]$. Assume that $P_{\textrm{pil}}$ pilots are utilized for joint estimation of channel ambiguities and IQ imbalance. By ZF estimation, the received signal vector $\tilde{\textbf{y}}_1$ is multiplied with the pseudoinverse of $\bar{\textbf{G}}$, obtaining $\textbf{r}=\bar{\textbf{G}}^{\dagger}\textbf{y}_1$. Define $\textbf{r}=[\textbf{r}_{\textrm{I},1}^T,\textbf{r}_{\textrm{Q},1}^T,\cdots,\textbf{r}_{\textrm{I},U}^T,\textbf{r}_{\textrm{Q},U}^T]^T$, where $\textbf{r}_{\textrm{I},u}$ and $\textbf{r}_{\textrm{Q},u}$ are respectively given by
\begin{equation}
\textbf{r}_{\textrm{I},u}=\bar{\textbf{d}}_{1,u}a_u+\textbf{z}_{\textrm{I},u}, \quad \textbf{r}_{\textrm{Q},u}=\bar{\textbf{d}}_{1,u}^*b_u+\textbf{z}_{\textrm{Q},u}
\end{equation}
with $\textbf{z}_{\textrm{I},u}$ and $\textbf{z}_{\textrm{Q},u}$ being the noise vectors. It is noteworthy that the ICI and ISI from the data symbols are eliminated thanks to the ZF algorithm, assuming the filter matrix $\textbf{A}$ is well-conditioned and its inverse exists \cite{6}. Hence, the following estimation of channel ambiguities and IQ imbalance parameters is more robust against ICI and ISI. By utilizing $P_{\textrm{pil}}$ pilots in the first GFDM symbol, the estimates of $a_u$ and $b_u$ in \eqref{eqbbb} are given by
\begin{equation}
\hat{a}_u=\frac{1}{P_{\textrm{pil}}}\sum\nolimits_{p=1}^{P_{\textrm{pil}}}\frac{\textbf{r}_{\textrm{I},u}(p)}{\bar{\textbf{d}}_{1,u}(p)}, \quad \hat{b}_u=\frac{1}{P_{\textrm{pil}}}\sum\nolimits_{p=1}^{P_{\textrm{pil}}}\frac{\textbf{r}_{\textrm{Q},u}(p)}{\bar{\textbf{d}}_{1,u}^*(p)}
\end{equation}

Define $\textbf{h}_{\textrm{I},u}$  and $\textbf{h}_{\textrm{Q},u}$ as the equivalent CIRs of user $u$ by taking into account IQ imbalance parameters, which are expressed as $\textbf{h}_{\textrm{I},u}=\textbf{h}_{u}\alpha_u$ and $\textbf{h}_{\textrm{Q},u}=\textbf{h}_{u}\beta_u$, respectively. The estimates of $\textbf{h}_{\textrm{I},u}$  and $\textbf{h}_{\textrm{Q},u}$ are given by 
\begin{equation}
\hat{\textbf{h}}_{\textrm{I},u}=\hat{\textbf{h}}_{u,0}\hat{a}_u, \quad \hat{\textbf{h}}_{\textrm{Q},u}=\hat{\textbf{h}}_{u,0}\hat{b}_u
\end{equation}
where $\hat{\textbf{h}}_{u,0}$ is the initial blind channel estimate obtained in Subsection \uppercase\expandafter{\romannumeral3}-B.

\subsection{Signal Detection Exploiting Diversity of IQ Imbalance}
Thanks to the diversity provided by IQ imbalance, signal can be detected with an enhanced accuracy in the following. Defining $\textbf{D}_u=\hat{\textbf{G}}_{\textrm{I},u}\hat{a}_u$ and $\textbf{F}_u=\hat{\textbf{G}}_{\textrm{Q},u}\hat{b}_u$, $\textbf{y}_i$ in \eqref{eq00001} can be rewritten as
\begin{equation}
\textbf{y}_i=\sum\nolimits_{u=1}^{U}\textbf{D}_u\textbf{d}_{i,u}+\sum\nolimits_{u=1}^{U}\textbf{F}_u\textbf{d}_{i,u}^*+\textbf{w}_i
\end{equation}
By concatenating $\textbf{y}_i$ and its conjugate $\textbf{y}_i^*$, we can obtain
\begin{equation}
\begin{bmatrix}\textbf{y}_i\\\textbf{y}_i^*\end{bmatrix}=\begin{bmatrix}\textbf{D} \quad \textbf{F}\\\textbf{F}^* \quad \textbf{D}^*\end{bmatrix}\begin{bmatrix}\textbf{d}_{i}\\\textbf{d}_{i}^*\end{bmatrix}+\begin{bmatrix}\textbf{w}_i\\\textbf{w}_i^*\end{bmatrix}
\end{equation}
where $\textbf{D}=[\textbf{D}_1,\cdots,\textbf{D}_{U}]$ of size $(G-L+1)N_{\textrm{r}}\times MK_{\textrm{D}}$, $\textbf{F}=[\textbf{F}_1,\cdots,\textbf{F}_{U}]$ of size $(G-L+1)N_{\textrm{r}}\times MK_{\textrm{D}}$ and $\textbf{d}_i=[\textbf{d}_{i,1}^T,\cdots,\textbf{d}_{i,U}^T]^T$ of size $MK_{\textrm{D}}\times 1$.

Denote $\tilde{\textbf{d}}_{\textrm{I},i}$ and $\tilde{\textbf{d}}_{\textrm{Q},i}$ as the respective estimates of the source signal $\textbf{d}_{i}$ and its image signal $\textbf{d}_{i}^*$, which can be obtained as
\begin{equation}
\begin{bmatrix}\hat{\textbf{d}}_{\textrm{I},i}\\\hat{\textbf{d}}_{\textrm{Q},i}\end{bmatrix}=\begin{bmatrix}\textbf{D} \quad \textbf{F}\\\textbf{F}^* \quad \textbf{D}^*\end{bmatrix}^{\dagger}\begin{bmatrix}\textbf{y}_i\\\textbf{y}_i^*\end{bmatrix}
\end{equation}
Since the source signal estimate $\hat{\textbf{d}}_{\textrm{I},i}$ is the conjugate of the image signal estimate $\hat{\textbf{d}}_{\textrm{Q},i}$, \emph{i.e.}, $\hat{\textbf{d}}_{\textrm{Q},i}=\hat{\textbf{d}}_{\textrm{I},i}^*$, the estimate of source signal $\textbf{d}_{i}$ can be obtained as 
$\hat{\textbf{d}}_{i}=(\hat{\textbf{d}}_{\textrm{I},i}+\hat{\textbf{d}}_{\textrm{Q},i}^*)/2$, 
by exploiting the diversity resulting from the IQ imbalance. Note that the proposed JCCIQE scheme is independent of the modulation type, while the method in \cite{23} works only with a constant modulus constellation. 


\section{Performance and Complexity Analysis}
\subsection{CRLB Analysis}
The CRLB for blind multi-CFO, multi-channel and multi-IQ imbalance estimation is derived in this subsection, which provides an analytical benchmark for the proposed JCCIQE scheme. CRLB is determined by taking a derivative of the received signal vector with respect to the unknown variables vector. Generally, the received signal samples from all the received antennas within a frame should be collected as a column vector \cite{44}. However, the dimension of this column vector is likely to be very high, and the inverse of such a huge matrix is computationally prohibitive. To simplify the computation, an approximate CRLB is derived in the following, inspired by \cite{42}.


By incorporating the IQ imbalance parameters $\alpha_u$ and $\beta_u$ into the original circulant channel matrix $\textbf{H}_{u}$, the equivalent circulant channel matrices are obtained as $\textbf{H}_{\textrm{I},u}=\alpha_u\textbf{H}_{u}$ and $\textbf{H}_{\textrm{Q},u}=\beta_u\textbf{H}_{u}$. $\textbf{H}_{\textrm{I},u}$ and $\textbf{H}_{\textrm{Q},u}$ are same as $\textbf{H}_u$ in \eqref{eq00001}, except that $\textbf{h}_u$ in \eqref{eq00001} is replaced by $\textbf{h}_{\textrm{I},u}$ and $\textbf{h}_{\textrm{Q},u}$, respectively. Thus, we obtain a system model equivalent to that given by \eqref{eq01}, where the received signal vector in the $i$-th symbol can be rewritten as
\begin{equation}
\textbf{y}_i=\sum\nolimits_{u=1}^{U}\textbf{H}_{\textrm{I},u}\textbf{E}(\phi_u)\bm{\Psi}_u\textbf{d}_{i,u}+\sum\nolimits_{u=1}^{U}\textbf{H}_{\textrm{Q},u}\textbf{E}(\phi_u)\bm{\Psi}_u^{*}\textbf{d}_{i,u}^*+\textbf{w}_i
\end{equation}

Since the signal vectors $\textbf{d}_{i,u}$ and $\textbf{d}_{i,u}^*$ are unknown to the receiver, the equivalent CIR vectors $\textbf{h}_{\textrm{I},u}$ and $\textbf{h}_{\textrm{Q},u}$ are estimated with a scaling ambiguity \cite{43}. Similar to \cite{43}, the last elements of the equivalent CIR vectors of user $u$ are assumed to be known at the receiver, which suggests that the channel ambiguities along with IQ imbalance are eliminated perfectly. Thus, the unknown variables are the CFO vector $\bm{\phi}=[\phi_1,\cdots,\phi_{U}]^T$, the equivalent CIR vectors
$\textbf{h}_{\textrm{I}}=[\textbf{h}_{\textrm{I},1}(1:N_{\textrm{r}}L-1)^T,\cdots,\textbf{h}_{\textrm{I},U}(1:N_{\textrm{r}}L-1)^T]^T$ and $\textbf{h}_{\textrm{Q}}=[\textbf{h}_{\textrm{Q},1}(1:N_{\textrm{r}}L-1)^T,\cdots,\textbf{h}_{\textrm{Q},U}(1:N_{\textrm{r}}L-1)^T]^T$,
and the unknown data vector $\textbf{d}_i$. We collect all the unknown variables in a column vector, $\bm{\Theta}=[\bm{\phi}^T,\textrm{Re}\{\textbf{h}^T_{\textrm{I}}\},\textrm{Im}\{\textbf{h}^T_{\textrm{I}}\},\textrm{Re}\{\textbf{h}^T_{\textrm{Q}}\},\textrm{Im}\{\textbf{h}^T_{\textrm{Q}}\},\textrm{Re}\{\textbf{d}_i^T]^T\},\textrm{Im}\{\textbf{d}_i^T]^T\}$.
Define $\textbf{V}_i=\sum\nolimits_{u=1}^{U}\textbf{H}_{\textrm{I},u}\textbf{E}(\phi_u)\bm{\Psi}_u\textbf{d}_{i,u}+\sum\nolimits_{u=1}^{U}\textbf{H}_{\textrm{Q},u}\textbf{E}(\phi_u)\bm{\Psi}_u^{*}\textbf{d}_{i,u}^*$. According to \cite{44}, the Fisher information matrix (FIM) can be expressed as
\begin{equation}
\bm{\Pi}_i=\frac{2}{\sigma^2}\textrm{Re}\Big [\frac{\partial\textbf{V}_i^H}{\partial \bm{\Theta}}\frac{\partial \textbf{V}_i}{\partial \bm{\Theta}^T}\Big ]
\label{eq11}
\end{equation}

The derivative of $\textbf{V}_i$ with respect to CFO of user $u$ $\phi_u$ is denoted as $\textbf{p}_{i,u}$, which is given by $\textbf{p}_{i,u}=\frac{2\pi}{K}(\textbf{H}_{\textrm{I},u}\textbf{D}\textbf{E}(\phi_u)\bm{\Psi}_u \textbf{d}_{i,u}+\textbf{H}_{\textrm{Q},u}\textbf{D}\textbf{E}(\phi_u)\bm{\Psi}_u^{*}\textbf{d}_{i,u}^*)$, with $\textbf{D}=\textrm{diag}\{0,\cdots,G-1\}$. Define $\textrm{Re}\{h_{\textrm{I},u}^{n_{\textrm{r}}}(l)\}$ as the real part of the equivalent CIR response of user $u$ $\textbf{h}_{\textrm{I},u}$ corresponding to its $l$-th path and $n_{\textrm{r}}$-th receive antenna. By taking the derivative of $\textbf{V}_i$ with respect to $\textrm{Re}\{h_{\textrm{I},u}^{n_{\textrm{r}}}(l)\}$, we can obtain $\textbf{q}_{i,u,n_{\textrm{r}},l}=\textrm{circshift}(\textbf{a}_l,n_{\textrm{r}})\textbf{E}(\phi_u)\bm{\Psi}_u \textbf{d}_{i,u}$, where $\textbf{a}_l=\textrm{circshift}(\textbf{a},l,2)$, and $\textbf{a}$ is the same as $\textbf{H}_u$ in \eqref{eq00001}, except that the elements with $h_{u}^0(L)$ in \eqref{eq00001} are replaced by ones and all the other elements are  by zeros. Thus, the derivative of $\textbf{V}_i$ with respect to the real part of the $u$-th equivalent CIR vector $\textrm{Re}\{\textbf{h}^T_{\textrm{I},u}\}$ is obtained as $\textbf{q}_{i,u}=\left[\textbf{q}_{i,u,1},\cdots,\textbf{q}_{i,u,N_{\textrm{r}}}\right]$, where $\textbf{q}_{i,u,n_{\textrm{r}}}=[\textbf{q}_{i,u,n_{\textrm{r}},1},\cdots,\textbf{q}_{i,u,n_{\textrm{r}},L}]$ for $n_{\textrm{r}}=1,\cdots,N_{\textrm{r}}-1$ and $\textbf{q}_{i,u,n_{\textrm{r}}}=[\textbf{q}_{i,u,N_{\textrm{r}}-1,1},\cdots,\textbf{q}_{i,u,N_{\textrm{r}}-1,L-1}]$ for $n_{\textrm{r}}=N_{\textrm{r}}$. Define $\textbf{t}_{i,u}=\textbf{H}_{\textrm{I},u}\textbf{E}(\phi_u)\bm{\Psi}_u+\textbf{H}_{\textrm{Q},u}\textbf{E}(\phi_u)\bm{\Psi}_u^*$ and $\textbf{u}_{i,u}=\textbf{H}_{\textrm{I},u}\textbf{E}(\phi_u)\bm{\Psi}_u-\textbf{H}_{\textrm{Q},u}\textbf{E}(\phi_u)\bm{\Psi}_u^*$ as the derivative of $\textbf{V}_i$ with respect to the real and imaginary parts of $\textbf{d}_{i,u}$, respectively. Hence, we can obtain
\begin{equation}
\frac{\partial\textbf{V}_i^H}{\partial \bm{\Theta}}=[\jmath\textbf{P}_i,\textbf{Q}_i,\jmath\textbf{Q}_i,\textbf{S}_i,\jmath\textbf{S}_i,\textbf{T}_i,\jmath\textbf{U}_i]^H, 
\end{equation}
and
\begin{equation}
\frac{\partial\textbf{V}_i}{\partial \bm{\Theta}^T}=[\jmath\textbf{P}_i,\textbf{Q}_i,\jmath\textbf{Q}_i,\textbf{S}_i,\jmath\textbf{S}_i,\textbf{T}_i,\jmath\textbf{U}_i]
\label{eq12}
\end{equation}
where $\textbf{P}_i=[\textbf{p}_{i,1},\cdots,\textbf{p}_{i,U}]$; $\textbf{Q}_i=[\textbf{q}_{i,1},\cdots,\textbf{q}_{i,U}]$; $\textbf{S}_i$ is obtained in a similar way to $\textbf{Q}_i$ but $\textbf{q}_{i,u,n_{\textrm{r}},l}=\textrm{circshift}(\textbf{a}_l,n_{\textrm{r}})\textbf{E}(\phi_u)\bm{\Psi}_u \textbf{d}_{i,u}$ is replaced by $\textbf{q}_{i,u,n_{\textrm{r}},l}=\textrm{circshift}(\textbf{a}_l,n_{\textrm{r}})\textbf{E}(\phi_u)\bm{\Psi}_u^* \textbf{d}_{i,u}^*$; $\textbf{T}_i=[\textbf{t}_{i,1},\cdots,\textbf{t}_{i,U}]$; and $\textbf{U}_i=[\textbf{u}_{i,1},\cdots,\textbf{u}_{i,U}]$.

According to \cite{42}, the approximate FIM is computed by $\bm{\Pi}_{\textrm{appro}}=\sum\nolimits_{i=1}^{N_{\textrm{s}}}\bm{\Pi}_i$.
Thus, the CRLB for blind estimation of multiple CFOs and equivalent channels can be obtained as the diagonal elements of $\bm{\chi}_{\textrm{appro}}$, where $\bm{\chi}_{\textrm{appro}}=\bm{\Pi}_{\textrm{appro}}^{-1}$. Since the CRLB on MSE of the equivalent channel estimation is derived assuming perfect estimation of channel ambiguities and IQ imbalance, it is unfair to compare it with the proposed JCCIQE scheme without perfect ambiguities and IQ imbalance estimation. Hence, similar to \cite{43}, we focus on the CRLB on MSE of CFO estimation, which is given by
\begin{equation}
\textrm{CRLB}_{\textrm{CFO}}=\frac{1}{U}\sum\nolimits_{v=1}^{U}\bm{\chi}_{\textrm{appro}}(v,v)
\end{equation}

\subsection{Complexity Analysis}
The complexity of the proposed JCCIQE scheme is analyzed, in comparison to the pilot aided estimation approaches \cite{15,9} for multiuser GFDM and the semi-blind joint multi-CFO and multi-channel estimation approach for OFDMA \cite{24}. Regarding the existing approaches for multiuser GFDM systems, the PN approach in \cite{15} is adopted to estimate multiple CFOs, while multiple IQ imbalances are estimated along with multiple channels by LS approach \cite{9}. Their symbolic computational complexities are demonstrated in Table \ref{tab1}, in terms of the number of complex additions and multiplications. They are compared in five aspects, namely multi-CFO estimation, multi-channel estimation, multi-channel ambiguities estimation, multi-IQ imbalance estimation and signal detection. The following observations can be made from Table \ref{tab1}.

\begin{table*}[htbp]
	\footnotesize
	\caption{Analytical computational complexity ($K$: Number of subcarriers per GFDM subsymbol/OFDMA symbol, $M$: Number of subsymbols per GFDM symbol, $U$: Number of users, $L$: Channel length, $P_{\textrm{pil}}$: Number of pilots, $N_{\textrm{r}}$: Number of receive antennas, $N_{\textrm{s}}$: Number of symbols in a frame, $N_u$: Number of total subcarriers assigned to user $u$, $\delta$: Search step size in the coarse CFO estimation, $\xi$: Number of cost function evaluations in the fine CFO estimation, $\Delta$: CFO search step size in \cite{24}, N/A: Not available).}
	
	\centering
	\begin{tabular}{|c|c|c|c|}
		\hline
		Item & JCCIQE& PN \cite{15}+LS \cite{9} & Zhang's scheme \cite{24} \\\hline
		\multirow{2}{*}{Multi-CFO estimation} & $K^2M^2N_{\textrm{r}}^2N_{\textrm{s}}+K^3M^3N_{\textrm{r}}^3$& \multirow{2}{*}{$KM/2+N^2N_{\textrm{s}}U$} & $U/\Delta(N_{\textrm{s}}N_{\textrm{r}}K(2K+\textrm{log}_2K)$ \\
		& $+QUN_{\textrm{r}}L(\frac{1}{\delta}+\xi)$ & & $+2N_{\textrm{r}}N_{\textrm{s}}(K-K/U)(K+N_{\textrm{r}})$ \\
		\cline{1-1}\cline{3-3}
		\multirow{2}{*}{Multi-channel estimation} & $(K^2M^2+KMN_u+N_{\textrm{r}}LN_u)$& $(8U^2L^2N+8U^3L^3$ & $+N_{\textrm{r}}^3)+U(2KN_{\textrm{s}}N_{\textrm{r}}+(N_{\textrm{r}}$\\
		& & $+2ULN)N_{\textrm{r}}$ & $-(U-1)L)^2)(N_{\textrm{r}}-(U-1)L)$ \\\hline
		Multi-channel ambiguity & $(N_{\textrm{r}}K^3M^3+2N_{\textrm{r}}N_uK^2M^2)U$& \multirow{4}{*}{N/A} & \multirow{2}{*}{$(2L+6)(N_{\textrm{r}}-(U-1)L)UP_{\textrm{pil}}$}\\
		estimation & $+16N_u^2N_{\textrm{r}}U^2KM$ & & \\
		\cline{1-1}\cline{4-4}
		Multi-IQ imbalance & $+2UP_{\textrm{pil}}$ & & \multirow{2}{*}{N/A}\\
		estimation & & & \\\hline
		Signal detection &\multicolumn{2}{c|}{$16N_u^2U^2KMN_{\textrm{r}}+4N_uUN_{\textrm{s}}KMN_{\textrm{r}}$} & $(N_{\textrm{r}}-(U-1)L)(2L+6)KN_{\textrm{s}}$ \\\hline
		
	\end{tabular}
	\label{tab1}%
\end{table*}%

First, the CFO and channel of each user are estimated jointly for the JCCIQE scheme and Zhang's scheme \cite{24}. Once the CFO is determined by minimizing the smallest eigenvalue, its corresponding channel can be easily obtained by the eigenvector, without the need to feed multiple CFOs back to multiple transmitters for multi-CFO compensation. In contrast, LS approach \cite{9} should be performed after multiple CFOs are estimated at the receiver and then compensated at each individual transmitter, leading to high training overhead and high latency.

Second, the PN \cite{15} and LS \cite{9} approaches are able to estimate multiple CFOs, multiple channels and multiple IQ imbalances with closed-form solutions, contributing to low computational complexity. On the contrary, the proposed JCCIQE scheme and Zhang's scheme \cite{24} suffer higher complexity due to the semi-blind implementation. For instance, the complexity of blind multi-CFO estimation depends on the CFO search step size, which is denoted by $\delta$ for JCCIQE and $\Delta$ for Zhang's scheme \cite{24}, respectively. Usually, there is a trade-off for the choice of the CFO search step size between the estimation accuracy and computational complexity. For example, a smaller step size contributes to a higher CFO estimation accuracy, while at the expense of a large number of computations. Thanks to the two-step CFO search of the proposed JCCIQE scheme, the coarse CFO search step size $\delta$ of JCCIQE can be chosen to be much larger than the CFO search step size $\Delta$ of Zhang's scheme \cite{24}. Regarding the fine CFO search of the proposed JCCIQE scheme, we count the number of cost function evaluations $\xi$ when implementing the golden section search and parabolic interpolation algorithms, which is shown to be less than 10 by simulation, making its complexity negligible compared to that of coarse CFO search.


Last but not least, the proposed JCCIQE scheme and Zhang's scheme \cite{24} are semi-blind, which requires very few pilots to resolve the channel ambiguities, while the PN \cite{15} and LS \cite{9} do not have the issue of ambiguities by estimating multiple CFOs, multiple channels and multiple IQ imbalances with a large number of pilots. Meanwhile, the proposed JCCIQE scheme is capable of estimating channel ambiguities and IQ imbalance jointly, while Zhang's scheme \cite{24} does not take into account IQ imbalance estimation.

\begin{table}[htbp]
	\footnotesize
	\caption{Normalized numerical computational complexity.}
	
	\centering
	\begin{tabular}{|c|c|c|c|c|}
		\hline
		System & \multicolumn{2}{c|}{Multiuser GFDM} & \multicolumn{2}{c|}{OFDMA} \\\hline
		Item & JCCIQE & PN \cite{15}+LS \cite{9} & JCCIQE & Zhang's scheme \cite{24} \\\hline
		Multi-CFO estimation & \multirow{2}{*}{$8.07\times 10^7$} & $5\times 10^4$ & \multirow{2}{*}{$8.07\times 10^7$} & \multirow{2}{*}{$8.62\times 10^8$} \\
		\cline{1-1}\cline{3-3}
		Multi-channel estimation &  & $3.5\times 10^3$ & &  \\\hline
		Multi-channel ambiguity estimation & \multirow{2}{*}{$7.58\times 10^5$} & \multirow{2}{*}{N/A} & \multirow{2}{*}{$7.58\times 10^5$} & 1 \\
		\cline{1-1}\cline{5-5}
		Multi-IQ imbalance estimation & & & & N/A \\\hline
		Signal detection & \multicolumn{2}{c|}{$3.8\times 10^5$} & $3.8\times 10^5$ & $6.4\times 10^3$ \\\hline
		Total & $8.07\times 10^7$ & $4.34\times 10^5$ & $8.07\times 10^7$ & $8.62\times 10^8$ \\\hline
	\end{tabular}
	\label{tab2}%
\end{table}%

The values of system parameters in Table \ref{tab1} are set as follows. Regarding multiuser GFDM systems, we set the number of subsymbols per GFDM symbol to $M=4$, the number of subcarriers per GFDM subsymbol to $K=16$, the number of users to $U=2$, the number of total subcarriers for each user to $N_u=28$, the channel length to $L=3$, the number of pilots for the estimation of multiple channel ambiguities and multiple IQ imbalances to $P_{\textrm{pil}}=1$, the number of receive antennas to $N_{\textrm{r}}=4$, the frame length to $N_{\textrm{s}}=200$, the search step size of JCCIQE to $\delta=0.01$ and the number of cost function evaluations to $\xi=6$. The parameters settings for OFDMA systems are the same with those for multiuser GFDM systems, except for $K$. Following IEEE 802.11ac standard \cite{bb4}, each OFDMA symbol contains $K=64$ subcarriers with $K_{\textrm{D}}=56$ data subcarriers. The CFO search step size for Zhang's scheme \cite{24} is $\Delta=0.001$. By substituting the values of those parameters into Table \ref{tab1}, the numerical complexity of the proposed JCCIQE scheme, PN \cite{15}, LS \cite{9} and Zhang's scheme \cite{24} are obtained, and then normalized to the lowest complexity of all items (\emph{i.e.}, complexity of multi-channel ambiguity estimation of Zhang's scheme \cite{24}), as shown in Table \ref{tab2}.

Two observations can be made from Table \ref{tab2}. On one hand, the PN and LS based approaches \cite{15,9} have the lowest complexity, thanks to the close-form solutions, while at the cost of high latency and high training overhead, as can be seen from Table \ref{tab3}. Moreover, the proposed JCCIQE scheme is much more computationally efficient than Zhang's scheme \cite{24}, with approximately 10-fold complexity reduction. The complexities of the proposed JCCIQE scheme for multiuser GFDM and OFDMA are comparable with each other. On the other hand, multi-CFO and multi-channel estimation dominates the overall complexity of JCCIQE and Zhang's \cite{24} schemes, due to the blind implementation. The complexities of the estimation of multiple channel ambiguities and multiple IQ imbalances and signal detection can be negligible.

\section{Simulation Results}
\subsection{Simulation Setup}
Monte Carlo simulations have been carried out to demonstrate the performance of the proposed JCCIQE scheme for multiuser GFDM systems in the presence of multiple CFOs and multiple IQ imbalances. As discussed in Subsection \uppercase\expandafter{\romannumeral4}-B, the pilot assisted approaches namely PN \cite{15} and LS \cite{9} are adopted for comparison with the proposed JCCIQE scheme in multiuser GFDM systems. Zhang's scheme \cite{24}, as a semi-blind approach, is also selected for performance comparison. The CRLB on multi-CFO estimation derived in Subsection \uppercase\expandafter{\romannumeral4}-A is exploited as a benchmark in Figs. \ref{Fig.3} and \ref{Fig.11}.

The setting of simulation parameters follows that in Subsection \uppercase\expandafter{\romannumeral4}-B, unless otherwise stated. Each GFDM symbol contains $M=4$ subsymbols, except for Figs. \ref{Fig.7} and \ref{Fig.4}. The number of users is $U=2$, except for Figs. \ref{Fig.2} and \ref{Fig.4}. The channel follows an exponential delay profile \cite{bb8} with channel length of $L=3$ and the normalized root mean square delay spread of 1.5. Assume channel exhibits Rayleigh fading. The CP length is $L_{\textrm{cp}}=4$. Each frame has a length of $N_{\textrm{s}}=200$ symbols, except for Fig. \ref{Fig.7}. The CFOs are randomly generated in the range of $[-|\phi_{\textrm{max}}|,|\phi_{\textrm{max}}|)$. Unless otherwise stated, $\phi_{\textrm{max}}$ is 0.5. Root raised cosine prototype filter with roll-off coefficient of 0.4 \cite{16} is utilized for GFDM. The amplitude and phase mismatches of each user are randomly chosen from $[0.8,1.2]$ and $[-15^{\circ},15^{\circ}]$ \cite{bb2}, respectively. Quadrature phase shift keying (QPSK) modulation is utilized.

It is noteworthy that multiple CFOs are estimated blindly by the proposed JCCIQE scheme, and only few pilots are utilized to jointly estimate multiple channel ambiguities and multiple IQ imbalances. The number of pilots for joint estimation of channel ambiguities and IQ imbalances for each user is set as $P_{\textrm{pil}}=1$, except for Fig. \ref{Fig.8}. Zhang's scheme \cite{24} adopts the same number of pilots as that of the proposed JCCIQE scheme. Regarding PN \cite{15} and LS \cite{9} for multiuser GFDM systems, two identical subsymbols are exploited to estimate each CFO, and one GFDM symbol is utilized to estimate multiple IQ imbalances and multiple channels jointly. The total number of pilots of PN \cite{15} and \cite{9} is 128 pilots, which is 64-fold more than that of the proposed JCCIQE scheme. The numbers of pilots utilized by the proposed JCCIQE scheme, the pilot aided approaches (PN \cite{15}+LS \cite{9}) and Zhang's scheme \cite{24} are summarized and compared in Table \ref{tab3}.

\begin{table}[htbp]
	\small
	\caption{Minimum required number of pilots utilized for joint estimation of multiple CFOs, multiple channels and multiple IQ imbalances ($U=2$, $K=16$ and $M=4$).}
	
	\centering
	\begin{tabular}{|c|c|c|}
		\hline
		JCCIQE & PN \cite{15}+LS \cite{9} & Zhang's scheme \cite{24} \\\hline
		$U$ & $2UK+KM$ & $U$ \\\hline
		2 & 128 & 2\\\hline
	\end{tabular}
	\label{tab3}%
\end{table}%

The MSE of multi-CFO estimation, and the MSE of the equivalent multi-channel estimation with IQ imbalances included are respectively defined as 
\begin{equation}
\textrm{MSE}_{\textrm{CFO}}=\frac{1}{U}\sum\nolimits_{u=1}^{U}\mathbb{E}\{(\hat{\phi}_u-\phi_u)^2\}
\end{equation}
and
\begin{equation}
\textrm{MSE}_{\textrm{Channel,IQ}}=\frac{1}{2UN_{\textrm{r}}L}\sum\nolimits_{u=1}^{U}\mathbb{E}\{(\parallel \hat{\textbf{h}}_{\textrm{I},u}-\textbf{h}_{\textrm{I},u}\parallel^2_{\textrm{F}}
+\parallel \hat{\textbf{h}}_{\textrm{Q},u}-\textbf{h}_{\textrm{Q},u}\parallel^2_{\textrm{F}})\}
\end{equation}

\begin{figure}[!t]
	\centering
	\includegraphics[width=9cm]{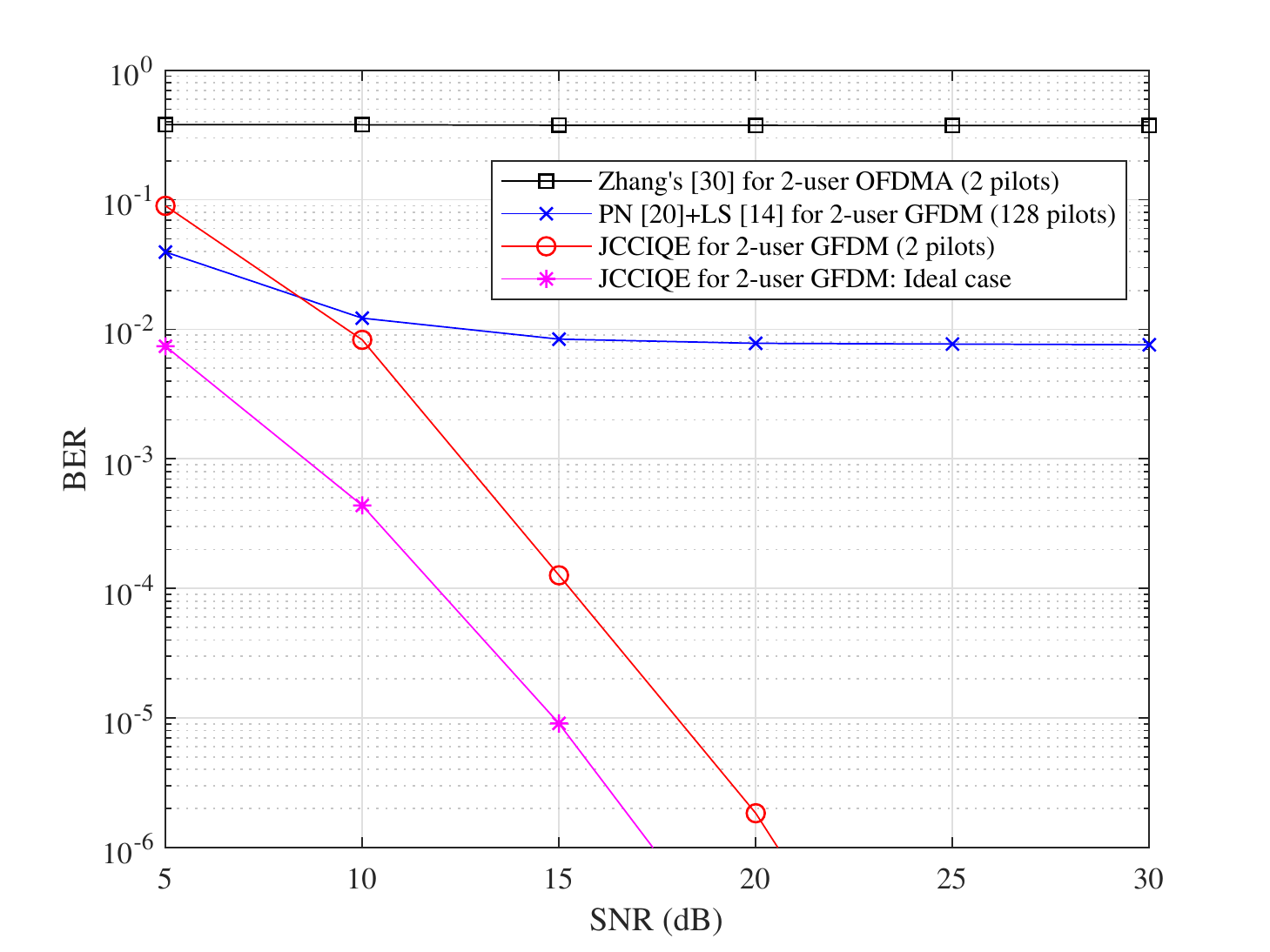}
	\caption{BER with $N_{\textrm{s}}=200$ GFDM symbols each with $M=4$ subsymbols, $N_{\textrm{r}}=4$ receive antennas;\protect\\
		Ideal case: perfect estimation of CFOs, channels and IQ imbalance.}
	\label{Fig.1}
\end{figure}

\subsection{BER Performance}

Fig. \ref{Fig.1} demonstrates the BER performance of the proposed JCCIQE scheme for 2-user GFDM, in comparison to Zhang's scheme for 2-user OFDMA \cite{24}, and PN \cite{15} and LS \cite{9} for 2-user GFDM system. The proposed JCCIQE scheme for 2-user GFDM outperforms Zhang's scheme \cite{24}, PN \cite{15} and LS \cite{9} from medium to high SNRs. PN \cite{15} and LS \cite{15} suffer an error floor at medium to high SNRs, due to the error propagation from multi-CFO estimation to the following estimation of multiple channels and IQ imbalances. Zhang's scheme \cite{24} demonstrates the worst performance at all SNRs. This is because Zhang's scheme \cite{24} requires at least $N_{\textrm{r}}=6$ receive antennas to work effectively with $U=2$ and $L=3$, while the proposed scheme only requires the number of receive antennas to be larger than or equal to $N_{\textrm{r}}=2$, which is independent of the number of users, as discussed in Subsection \uppercase\expandafter{\romannumeral3}-A. Note that the BER performance of the proposed JCCIQE scheme is slightly worse than that of PN \cite{15} and LS \cite{9} at $\textrm{SNR}=5$ dB. This is because JCCIQE is semi-blind and based on the second-order statistics of the received signal, whose robustness against noise can be enhanced by utilizing more receive antennas and more received symbols for blind estimation, as will be seen in Figs. \ref{Fig.6} and \ref{Fig.7}.

\begin{figure}[!t]
	\centering
	\includegraphics[width=9cm]{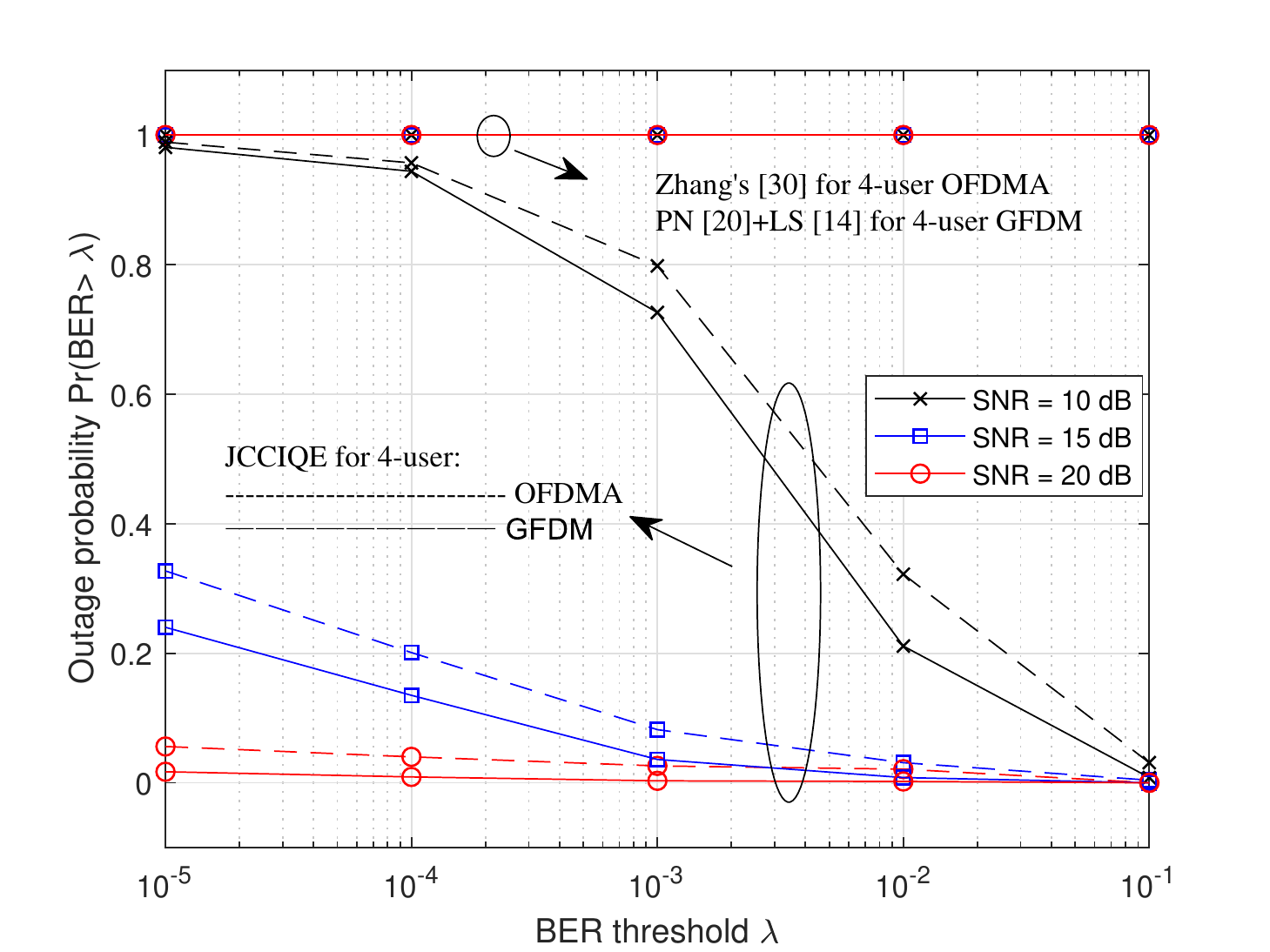}
	\caption{Outage probability with $U=4$ users, $N_{\textrm{r}}=4$ receive antennas, $N_{\textrm{s}}=200$ GFDM symbols each with $M=4$ subsymbols.}
	\label{Fig.2}
\end{figure}

Fig. \ref{Fig.2} shows the outage probabilities of the proposed JCCIQE scheme, Zhang's scheme \cite{24}, and PN\cite{15} and LS \cite{9}, with $U=4$ users, $M=4$ and $N_{\textrm{r}}=4$ receive antennas, at SNR $=10$, $15$ and $20$ dB. The outage probability is defined as the probability of the system BER being larger than a threshold $\lambda$. JCCIQE demonstrates a much lower outage probability than that of Zhang's scheme \cite{24}, and PN \cite{15} and LS \cite{9}, which have an outage probability of one. The high outage probability of Zhang's scheme \cite{24} is due to that it demands at least $N_{\textrm{r}}=12$ receive antennas with $U=4$ and $L=3$, while the proposed scheme only requires the number of receive antennas to be no less than $N_{\textrm{r}}=2$, same as that in Fig. \ref{Fig.1}. Moreover, PN \cite{15} is not working for GFDM with $U=4$ users and $M=4$ subsymbols, as the number of required subsymbols has to be at least $M=2U$, \emph{e.g.}, $M=8$ for $U=4$. While the proposed JCCIQE scheme can work with any value of $M$. Moreover, we can observe the outage probability of JCCIQE scheme for multiuser GFDM is slightly lower than that of JCCIQE for OFDMA systems.

\subsection{MSE of Equivalent Channel Estimation}

\begin{figure}[!t]
	\centering
	\includegraphics[width=9cm]{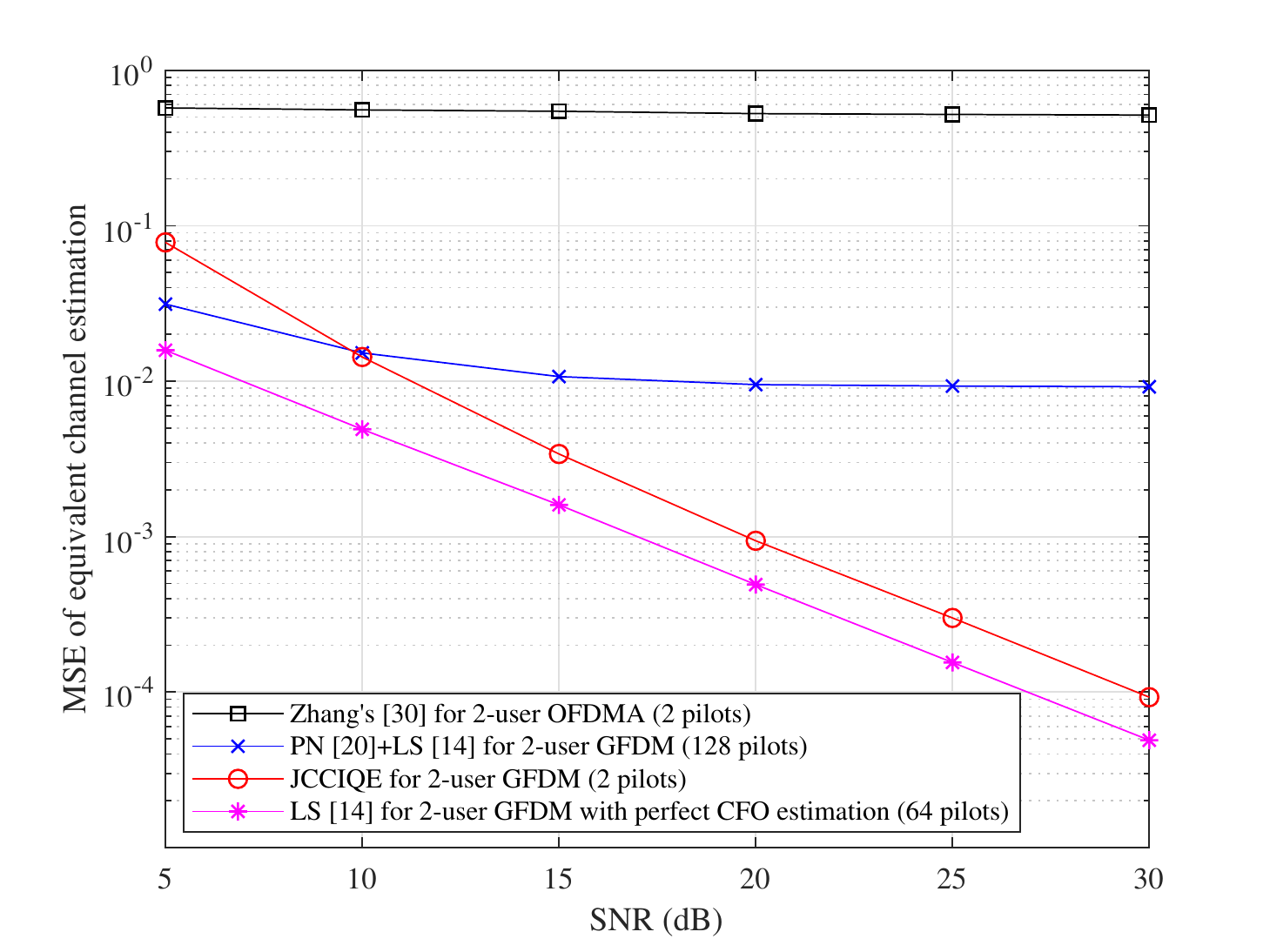}
	\caption{MSE of equivalent channel estimation, with $N_{\textrm{s}}=200$ GFDM symbols each with $M=4$ subsymbols and $N_{\textrm{r}}=4$ receive antennas.}
	\label{Fig.5}
\end{figure}

Fig. \ref{Fig.5} shows the MSE of equivalent channel estimation of the proposed JCCIQE scheme, in comparison to the LS \cite{9} with perfect multi-CFO estimation, LS \cite{9} with PN \cite{15}, and Zhang's scheme \cite{24}. Due to the error propagation from multi-CFO estimation, LS \cite{9} with PN \cite{15} suffers an error floor at medium to high SNRs. Zhang's scheme \cite{24} has the worst MSE of channel estimation performance, as it demands more receive antennas. Moreover, the MSE of channel estimation performance of the proposed JCCIQE scheme approaches that of LS \cite{9} with perfect multi-CFO estimation especially from SNR $=15$ dB to SNR $=30$ dB. Similarly to Fig. \ref{Fig.1}, JCCIQE exhibits a slightly worse MSE of channel estimation than that of LS \cite{9} with PN \cite{15} at SNR $=5$ dB.

\begin{figure}[!t]
	\centering
	\includegraphics[width=9cm]{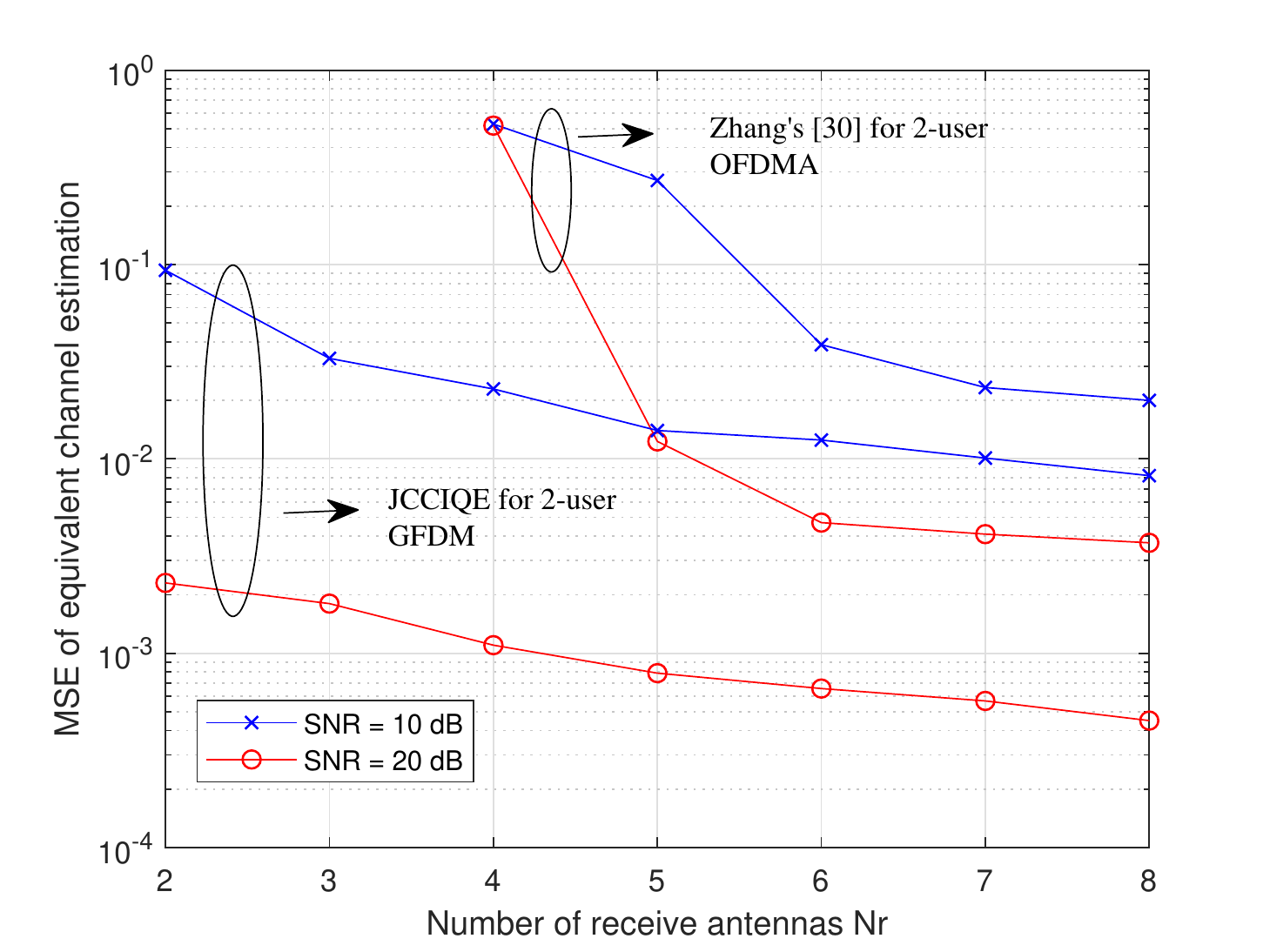}
	\caption{Impacts of the number of receive antennas on MSE of equivalent channel estimation, with $U=2$ users, SNR $=10$ dB and SNR $=20$ dB.}
	\label{Fig.6}
\end{figure}

Fig. \ref{Fig.6} investigates the impact of the number of receive antennas on the MSE of channel estimation performance of the proposed JCCIQE scheme and Zhang's scheme \cite{24}, with SNR $=10$ dB and SNR $=20$ dB. The MSE of channel estimation performance of Zhang's scheme \cite{24} is very poor at $N_{\textrm{r}}=4$. Its performance enhances as the number of receive antennas increases and converges when $N_{\textrm{r}}$ is 6. In contrast, the proposed JCCIQE scheme is able to provide a good channel estimation performance with $N_{\textrm{r}}=2$ for SNR $=20$ dB.

\begin{figure}[!t]
	\centering
	\includegraphics[width=9cm]{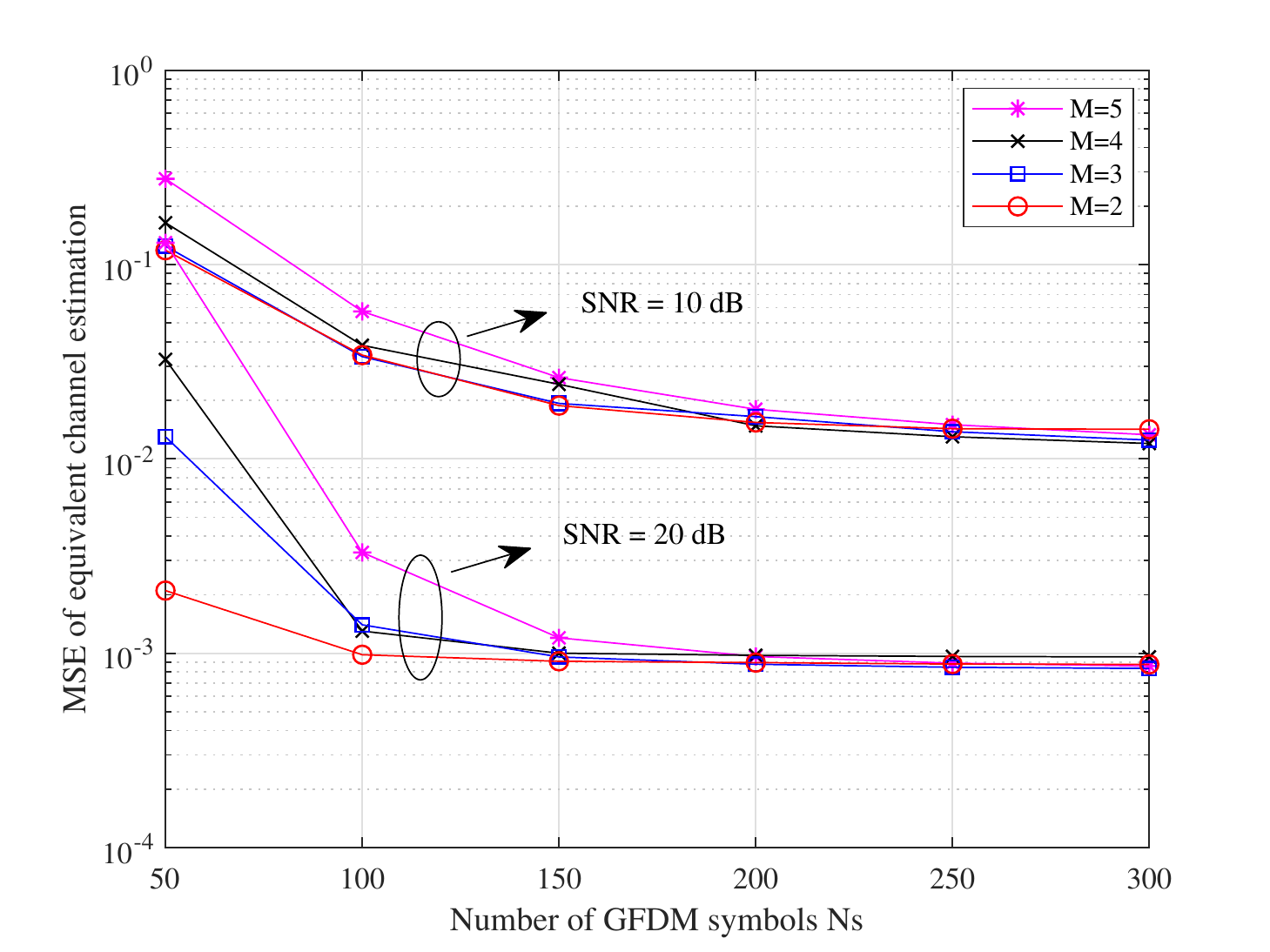}
	\caption{Impact of the number of subsymbols per GFDM symbol and the number of GFDM symbols on MSE of equivalent channel estimation, with $U=2$ users and $N_{\textrm{r}}=4$ receive antennas.}
	\label{Fig.7}
\end{figure}

Fig. \ref{Fig.7} exhibits the impacts of the number of subsymbols per GFDM symbol and the number of GFDM symbols on MSE of equivalent channel estimation of the proposed JCCIQE scheme, at SNR $=10$ and $20$ dB. For all values of $M$, utilizing more GFDM symbols lowers the MSE of channel estimation substantially. Given the number of GFDM symbols, the performance of the proposed JCCIQE scheme deteriorates as the value of $M$ increases. This is because the size of GFDM symbol increases with $M$, which in turn requires more GFDM symbols to achieve a good second-order statistics of the received signal for blind multi-CFO and multi-channel estimation.


\begin{figure}[!t]
	\centering
	\includegraphics[width=9cm]{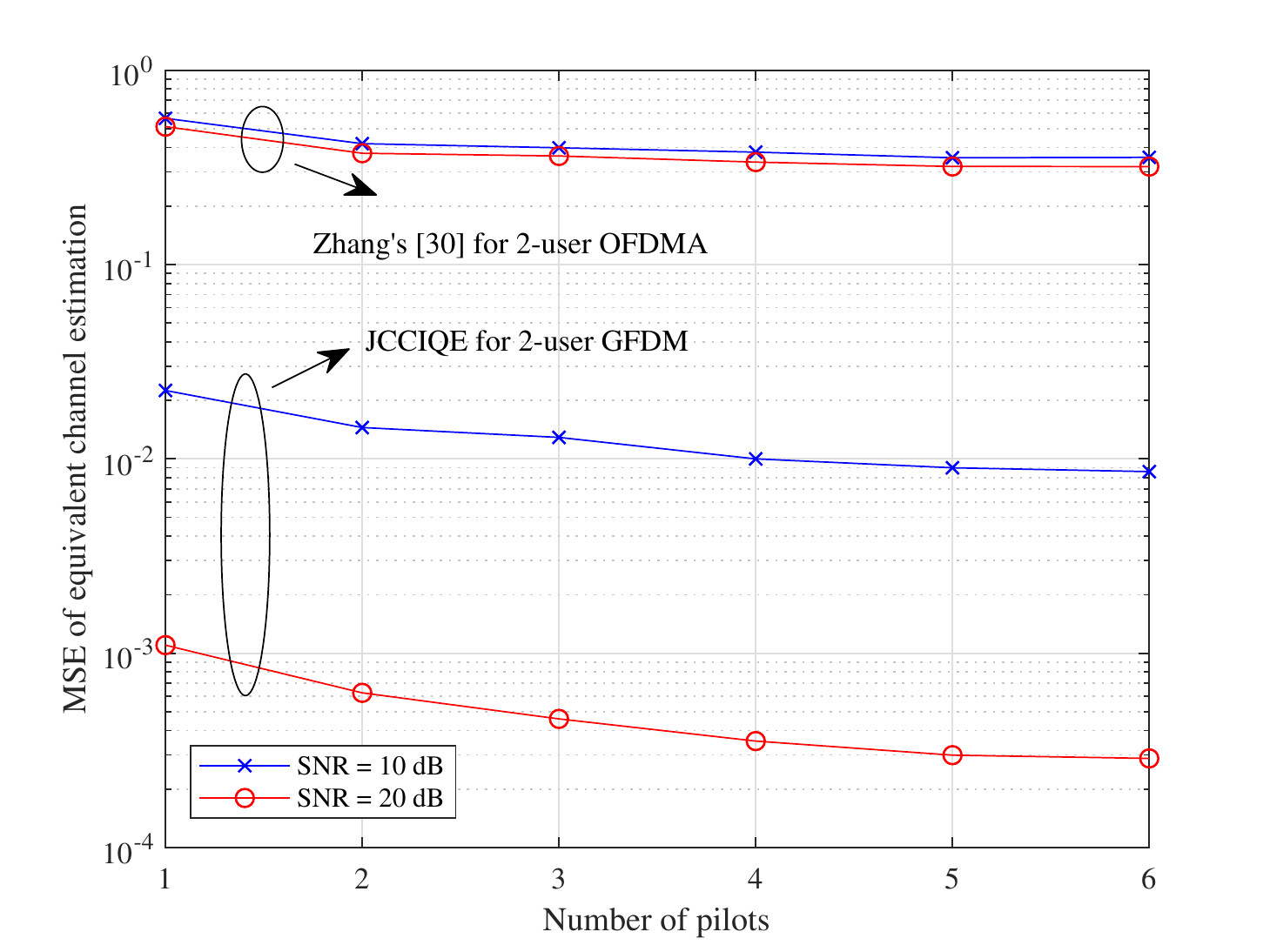}
	\caption{Impact of the number of pilots of each user on MSE of equivalent channel estimation, with $U=2$ users, $N_{\textrm{s}}=200$ GFDM symbols each with $M=4$ subsymbols, and $N_{\textrm{r}}=4$ receive antennas.}
	\label{Fig.8}
\end{figure}

Fig. \ref{Fig.8} demonstrates the impact of the number of pilots on MSE of equivalent channel estimation of the proposed JCCIQE scheme, in comparison to Zhang's scheme \cite{24} for OFDMA, at SNR $=10$ and $20$ dB. JCCIQE significantly outperforms Zhang's scheme \cite{24} with perfect estimation and compensation of IQ imbalances. Its performance improves with pilot length and reaches a steady state at $P_{\textrm{pil}}=4$ pilots.

\subsection{MSE of CFO Estimation}

\begin{figure}[!t]
	\centering
	\includegraphics[width=9cm]{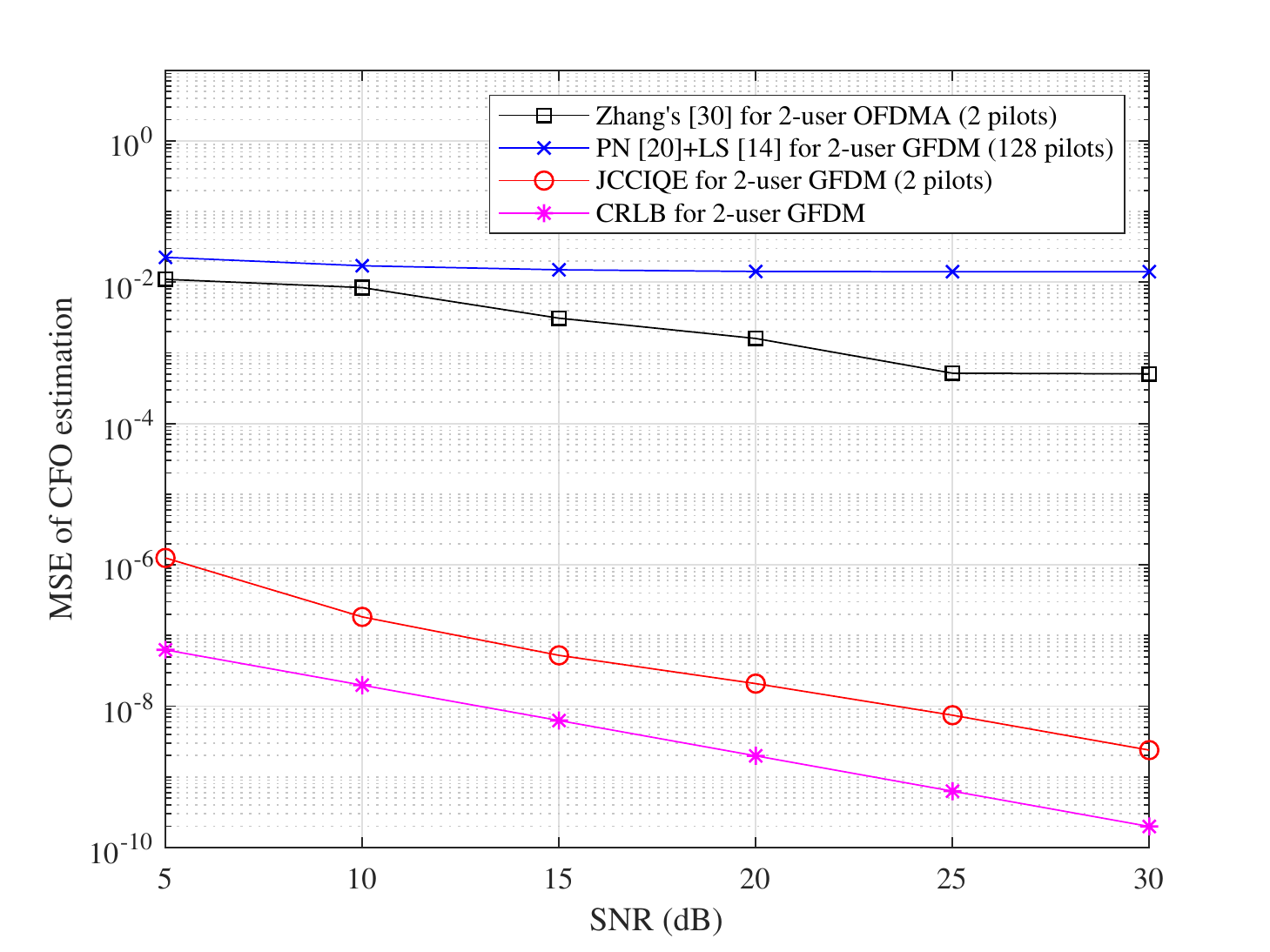}
	\caption{MSE of CFO estimation, with $N_{\textrm{s}}=200$ GFDM symbols each with $M=4$ subsymbols and $N_{\textrm{r}}=4$ receive antennas.}
	\label{Fig.3}
\end{figure}

\begin{figure}[!t]
	\centering
	\includegraphics[width=9cm]{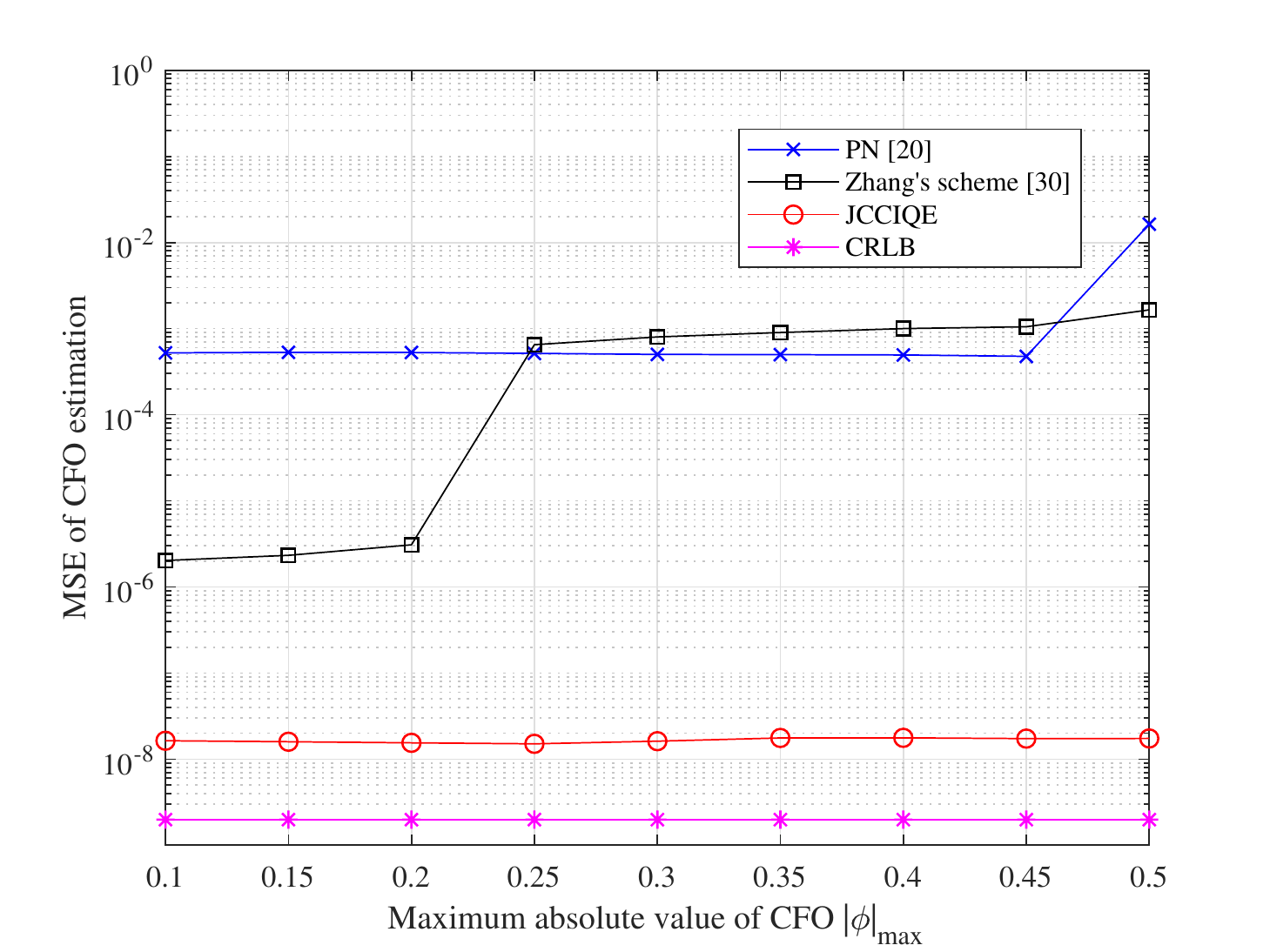}
	\caption{Impact of CFO on MSE of CFO estimation at SNR $=20$ dB, with $N_{\textrm{r}}=4$ receive antennas and $N_{\textrm{s}}=200$ symbols.}
	\label{Fig.11}
\end{figure}

Fig. \ref{Fig.3} shows the MSE of CFO estimation of the proposed JCCIQE scheme, in comparison to Zhang's scheme \cite{24} and PN \cite{15}. The proposed JCCIQE scheme demonstrates a substantially lower MSE performance than the the existing approaches \cite{15,24}. PN \cite{15} suffers an error floor at all SNRs, due to ICI, ISI and MUI. In contrast, JCCIQE separates and estimates multiple CFOs by subspace, without suffering ICI, ISI and MUI, as discussed in Section III. The MSE of CFO estimation by the proposed JCCIQE scheme is also close to the CRLB. As can be seen in Fig. \ref{Fig.11}, the proposed JCCIQE scheme is also more robust against CFO than the existing PN scheme \cite{15} and Zhang's scheme \cite{24}.

\begin{figure}[!t]
	\centering
	\includegraphics[width=9cm]{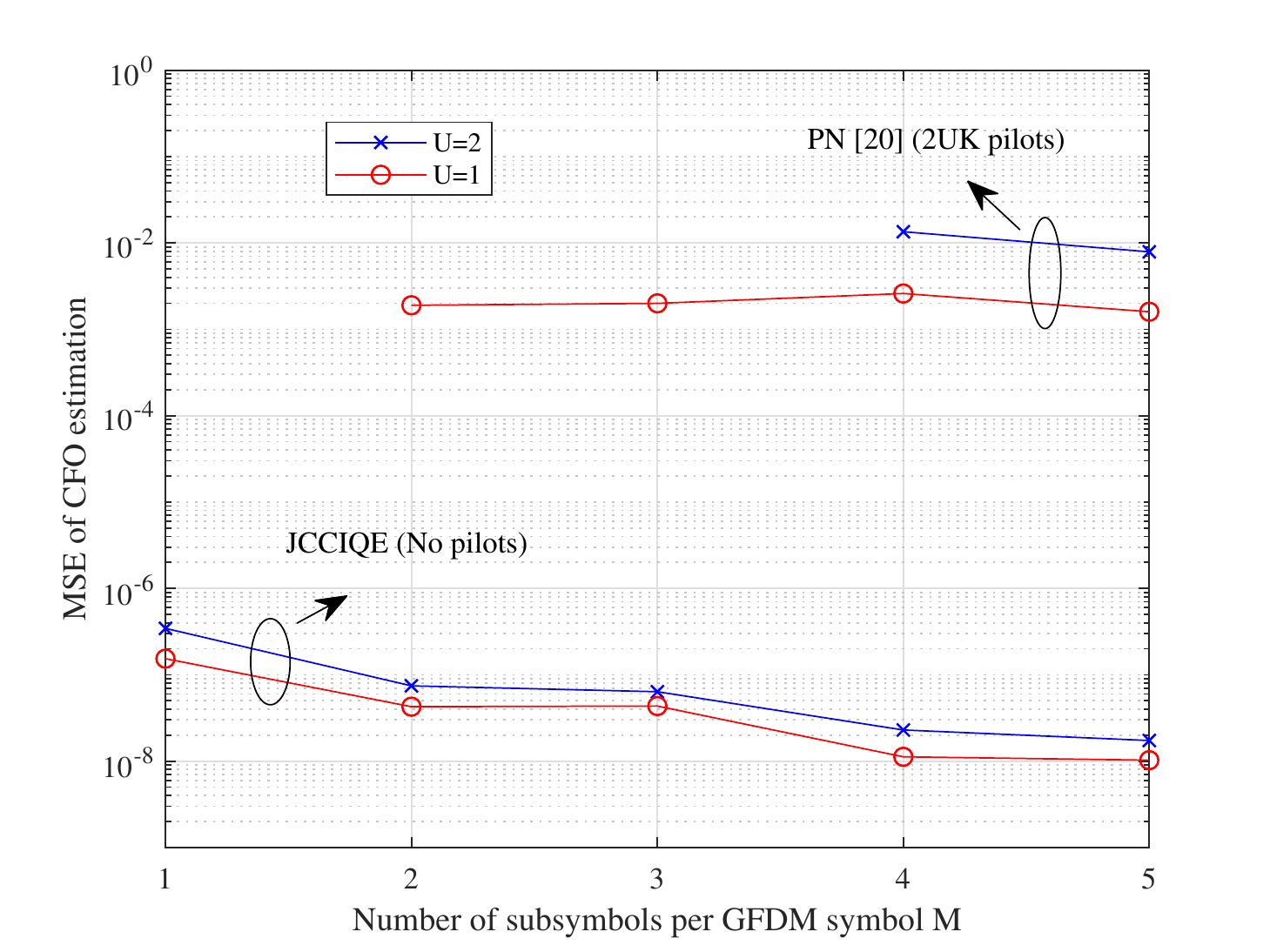}
	\caption{Impact of the number of subsymbols per GFDM symbol on MSE of CFO estimation at SNR $=20$ dB, with $N_{\textrm{r}}=4$ receive antennas and $N_{\textrm{s}}=200$ symbols.}
	\label{Fig.4}
\end{figure}

Fig. \ref{Fig.4} demonstrates the impact of the number of subsymbols per GFDM symbol on MSE of CFO estimation of the proposed JCCIQE scheme with $U=1$ and $U=2$, in comparison to PN \cite{15} at SNR $=20$ dB. The CFO estimation performance of the proposed JCCIQE scheme slightly enhances with the number of subsymbols $M$ regardless of the number of users, as more signal samples can be utilized for multi-CFO estimation. Meanwhile, the proposed JCCIQE scheme can work with any value of $M$, which is independent of the number of users. In contrast, the minimum required value of $M$ for PN based approach \cite{15} is $2U$, \emph{e.g.}, $M=2$ and $M=4$ for $U=1$ and $U=2$, respectively. Besides, the training overheads of PN \cite{15} increases with the number of users, while the proposed JCCIQE scheme is able to estimate multiple CFOs blindly without requiring any pilots, giving rise to high spectrum efficiency.

\begin{table*}[htbp]
	\footnotesize
	\caption{Comparison of the proposed JCCIQE scheme and the existing approaches.}
	\centering
	\begin{tabular}{|c|c|c|c|c|}
		\hline
		\multicolumn{1}{|c|}{Item} & JCCIQE    & PN \cite{15}  & LS \cite{9} & Zhang's scheme \cite{24} \\\hline
		Modulation & GFDM & GFDM & GFDM & OFDM \\\hline
		CFO estimation & Yes & Yes  & No  & Yes  \\\hline
		Channel estimation & Yes &  No  & Yes  & Yes  \\\hline
		IQ imbalance estimation & Yes &  No  & Yes  &  No  \\\hline
		Training overhead & Very low & Very high ($2UK$) & Very high ($KM$) & Very low \\\hline
		Number of required receive antennas & $\geq 2$ & $\geq 1$ & $\geq 1$ & $\geq UL$  \\\hline
		Number of required subsymbols & $\geq 1$ & $\geq 2U$ & $\geq 1$ & N/A \\\hline
		Robustness against ICI, ISI and MUI & High & Low & Low & N/A \\\hline
		Computational complexity & Medium & Low & Low & High \\\hline
		Robustness against CFO & High & Low & Low & Low \\\hline
	\end{tabular}
	\label{tab4}%
\end{table*}%

\section{Conclusion}
A low-complexity semi-blind estimation scheme has been proposed for a comprehensive multiuser GFDM system, taking into account multiple performance limiting factors at the same time, namely, CFOs, IQ imbalances, channel estimation, training overhead, ICI and ISI, as summarized in Table \ref{tab4} against other approaches \cite{15,9,24}. The proposed JCCIQE scheme has much lower training overhead, which is approximately $64$-fold lower than that of the existing pilot assisted approaches \cite{15,9}. Besides, no signaling feedback is demanded thanks to multi-CFO compensation performed at receiver side, unlike \cite{37}. JCCIQE demonstrates much superior performances in terms of BER, outage probability, MSE of CFO estimation and MSE of equivalent channel estimation than the existing methods \cite{15,9,24}. The performance of JCCIQE approaches the CRLB on MSE of CFO estimation. In addition, it is robust against CAS and does not suffer any error floors caused by the ICI and ISI encountered in the previous pilot assisted approaches \cite{15,9}. It can work with a small number of receive antennas and a smaller number of subsymbols per GFDM symbol, and has a complexity reduction of tens of times over Zhang's scheme \cite{24} for OFDMA. It is more robust against CFO than the existing PN \cite{15} and Zhang's \cite{24} scheme. This work will be extended to multiuser GFDM systems in the presence of multiple frequency dependent (FD) IQ imbalances, by deriving an equivalent channel model with FD IQ imbalances included, as in \cite{bb6} and \cite{bb7}.


\bibliographystyle{IEEEtran}
{\small
	\bibliography{IEEE_TWC}}

\IEEEpeerreviewmaketitle




\end{document}